\documentclass[aps,prx,amsmath,floats,floatfix,twocolumn,
  superscriptaddress,nofootinbib,showpacs]{revtex4-1}

\pdfoutput=1 
\usepackage{diagbox}
\usepackage{enumitem}
\usepackage{multirow}
\usepackage{braket}
\usepackage{amsfonts}
\usepackage{amsmath}
\usepackage{amssymb}
\usepackage{amsthm}
\usepackage{bm}
\usepackage{graphicx}
\usepackage{graphics}
\usepackage[latin1]{inputenc}
\usepackage{latexsym}
\usepackage{rotating}
\usepackage[dvipsnames]{xcolor}
\usepackage{mathrsfs}
\usepackage{microtype}
\usepackage{verbatim}
\usepackage{url}

\usepackage[caption=false]{subfig}
\usepackage[normalem]{ulem}

\usepackage[breaklinks=true]{hyperref}
\hypersetup{
    colorlinks=true,
    linkcolor=NavyBlue,
    filecolor=Magenta,
    urlcolor=NavyBlue,
    citecolor=NavyBlue
}

\usepackage{yfonts}
\usepackage{float}
\usepackage{xspace} 
\usepackage{mathrsfs}
\usepackage{siunitx}
\usepackage{array}

\newcommand{\be}{\begin{equation}}
\newcommand{\ee}{\end{equation}}

\newcommand{\ba}{\begin{align}}
\newcommand{\ea}{\end{align}}

\makeatletter
\newcommand*{\rom}[1]{\expandafter\@slowromancap\romannumeral #1@}
\makeatother

\AtBeginDocument{%
    \newwrite\bibnotes
    \def\bibnotesext{Notes.bib}
    \immediate\openout\bibnotes=\jobname\bibnotesext
    \immediate\write\bibnotes{@CONTROL{REVTEX41Control}}
    \immediate\write\bibnotes{@CONTROL{%
    apsrev41Control,author="08",editor="1",pages="1",title="0",year="1"}}
    \if@filesw
    \immediate\write\@auxout{\string\citation{apsrev41Control}}%
    \fi
}

\allowdisplaybreaks

\begin{document}

\preprint{APS/123-QED}

\title{Untargeted Bayesian search of anisotropic gravitational-wave backgrounds \\ through the analytical marginalization of the posterior}

\author{Adrian Ka-Wai Chung}
\email{akwchung@illinois.edu}
\affiliation{Illinois Center for Advanced Studies of the Universe \& Department of Physics, University of Illinois at Urbana-Champaign, Urbana, Illinois 61801, USA}

\author{Nicol\'as Yunes}
\affiliation{Illinois Center for Advanced Studies of the Universe \& Department of Physics, University of Illinois at Urbana-Champaign, Urbana, Illinois 61801, USA}

\date{\today}

\begin{abstract}
We develop a method to perform an untargeted Bayesian search for anisotropic gravitational-wave backgrounds that can efficiently and accurately reconstruct the background intensity map. 
Our method employs an analytic marginalization of the posterior of the spherical-harmonic components of the intensity map, without assuming the background possesses any specific angular structure. 
The key idea is that the likelihood function of the spherical-harmonic components is a multivariate Gaussian when the intensity map is expressed as a linear combination of the spherical-harmonic components and the noise is stationary and Gaussian. 
If a uniform and wide prior of these spherical-harmonic components is prescribed, the marginalized posterior and the Bayes factor can be well approximated by a high-dimensional Gaussian integral.
The analytical marginalization allows us to regard the spherical-harmonic components of the intensity map of the background as free parameters, and to construct their individual marginalized posterior distribution in a reasonable time, even though many spherical-harmonic components are required. 
The marginalized posteriors can, in turn, be used to accurately construct the intensity map of the background. 
By applying our method to mock data, we show that we can recover precisely the angular structures of various simulated anisotropic backgrounds, without assuming prior knowledge of the relation between the spherical-harmonic components predicted by a given model. 
Our method allows us to bypass the time-consuming numerical sampling of a high-dimensional posterior, leading to a more model-independent and untargeted Bayesian measurement of the angular structures of the gravitational-wave background. 
\end{abstract}

\maketitle

\allowdisplaybreaks[4] 

\section{Introduction}
\label{sec:Intro}

The direct detection of gravitational waves (GWs) emitted by compact binary coalescence (CBC) is a milestone in GW astrophysics \cite{LIGO_01, LIGO_02, LIGO_03, LIGO_04, LIGO_05, LIGO_06, LIGO_07, LIGO_08, LIGO_09, LIGO_10}. 
The detection of a GW background (GWB), formed by the random and incoherent superposition of numerous individually unresolvable GW signals emitted by different types of sources, may be the milestone that can be achieved next, in the foreseeable future \cite{LIGO_SGWB_01, LIGO_SGWB_02, LIGO_SGWB_03, LIGO_SGWB_04, Renzini:2022alw}. 
The North American Nanohertz Observatory for Gravitational Waves (NANOGrav) collaboration has just published $\sim 4\sigma$ significance evidence of the detection of a GWB by analysing its 15-year data set \cite{NANOGrav:2023gor, NANOGrav:2023ctt, NANOGrav:2023hde, NANOGrav:2023icp}. 
This discovery immediately opens up new directions of astronomical research \cite{NANOGrav:2023hfp, NANOGrav:2023hvm}. 
Astrophysical sources, including CBCs \cite{Zhu:2012xw, Zhao:2020iew, 10.1111/j.1365-2966.2003.07176.x, Capurri:2021zli}, rapidly rotating asymmetric neutron stars \cite{SGWB_RNS_01,SGWB_RNS_02, SGWB_RNS_03, SGWB_RNS_04, SGWB_RNS_05} and core-collapse supernova \cite{SGWB_CCSN_01, SGWB_CCSN_02, SGWB_CCSN_03, SGWB_CCSN_04}, can generate GWs that form a GWB. 
Alternatively, a GWB can also be generated by GWs emitted by cosmological sources, like cosmological inflation \cite{Caprini:2018mtu, Guzzetti:2016mkm, Contaldi:2018akz, Easther:2006vd, 1979JETPL..30..682S, Cook:2011hg, Turner:1996ck, Easther:2006gt}, the phase transitions that may have occurred in the early Universe  \cite{Romero:2021kby, 10.21468/SciPostPhysLectNotes.24, Weir:2017wfa, Caprini:2015zlo, Kahniashvili:2008pe, Kahniashvili:2009mf, Caprini:2009yp, Kisslinger:2015hua, RoperPol:2019wvy, Boileau:2022ter}, and cosmic strings \cite{Mairi_CS_SGWB_01, Mairi_CS_SGWB_02, Mairi_CS_SGWB_03, SGWB_CS_01, SGWB_CS_02, LIGOScientific:2017ikf, Jenkins:2018nty, LIGOScientific:2021nrg, SGWB_CS_03}, if they exist. 
A GWB may even be generated by physics that has yet to be fully explored, such as ultralight bosons \cite{SGWB_ULB_01, SGWB_ULB_02, SGWB_ULB_03, SGWB_ULB_04, SGWB_ULB_05} and primordial black holes \cite{SGWB_PBH_01, SGWB_PBH_02, SGWB_PBH_03, SGWB_PBH_04}, which are candidates to explain dark matter. 
As a GWB can be formed by sources significantly different from those generating individually detectable GW signals, detecting a GWB constitutes a unique probe of the Universe~ \cite{Renzini:2022alw}.

While a GWB is expected to be dominantly isotropic, it should also contain angular structures. 
In general, different sources and generation mechanisms could form GWBs with different angular structures \cite{Cusin:2017fwz, Jenkins:2018nty, Geller:2018mwu, Jenkins:2018uac, Jenkins:2018kxc, Bartolo:2019oiq, Cusin:2018rsq, Capurri:2021zli, Dimastrogiovanni:2019bfl, Dimastrogiovanni:2022eir}. 
This source and mechanism dependence suggests that accurately mapping the angular structure of the GWB could be very informative, allowing us to pinpoint GWB sources and deduce their properties \cite{Zheng:2023ezi}. 
To this end, several methods have been developed to extract the angular distribution of the GWB power spectrum. 
Broadly speaking, these methods can be classified as either \textit{frequentist} or \textit{Bayesian}. 
The frequentist approach amounts to constructing some maximum-likelihood estimator with different basis to characterize GWB anisotropies. 
Examples of the frequentist approach include radiometer search \cite{Ballmer:2005uw} and spherical-harmonic decomposition \cite{Thrane:2009fp}, which have been widely used in analyzing the actual data measured by the LIGO and Virgo detectors \cite{KAGRA:2021mth}. 
The Bayesian approach amounts to constructing the posterior of random variables related to GWB anisotropies, such as done very recently in~\cite{Tsukada:2022nsu, 2021MNRAS.507.5451B}. 

These two approaches are useful in probing GWB anisotropies, but they also have limitations. 
For example, since the radiometer search works in the pixel basis, it is not suitable for searching for extended sources \cite{LIGOScientific:2011kvu, LIGOScientific:2016nwa}. 
Working in the spherical-harmonic basis, one can use a spherical-harmonic decomposition to search for widespread sources and probe the anisotropies in a model-independent way, but it may lead to some nonphysical maximum likelihood estimates, such as complex estimates for some coefficients that, on physical grounds, should be real. 

One way to remedy the drawback of the spherical-harmonic decomposition is to perform a model-independent Bayesian search for an anisotropic GWB. 
However, to describe an anisotropic GWB without assuming any source models, one needs a model-independent framework that is typically characterized by many parameters, such as (the formally infinite number of) the spherical-harmonic components required in the spherical-harmonic decomposition. 
This large number of variables makes the construction of the posterior of these variables computationally untenable, even if the posterior is estimated through numerical sampling~\cite{2019PASA...36...10T}. 
Thus, the Bayesian search of an anisotropic GWB has been restricted to either a \textit{targeted} Bayesian analysis \cite{Tsukada:2022nsu}, inferring the overall amplitude of a GWB whose angular structures are given by a specific model, or to a model-independent framework characterized by only a few parameters \cite{2021MNRAS.507.5451B}. 

The goal of this paper is to develop a computationally efficient, fast and \textit{untargeted} Bayesian search that can construct the Bayesian marginalized posterior of the spherical-harmonic components of the angular structure of the anisotropic GWB without prior knowledge of the relation among the spherical harmonic components. 
We start with a spherical-harmonic decomposition of the intensity map of the GWB. 
 This decomposition allows us to express the energy flux of the GWB as a function of the sky direction through linear combinations of the spherical harmonic components. If the noise is stationary and Gaussian, then the likelihood function of the spherical-harmonic components is a multivariate Gaussian function.
Thus, the marginalized posterior of a specific spherica- harmonic component and the Bayes factor (between an anisotropic GWB and a nondetection hypotheses) can be well approximated by a high-dimensional Gaussian integral. 
After evaluating this integral, the marginalized posterior of the real or imaginary part of a particular spherical-harmonic component is also a Gaussian function, whose mean and variance are given by the convolution between the cross-spectral density of the data  (i.e.~the product of the frequency-domain data measured by a detector and the complex conjugate of the frequency-domain data measured by another detector, see Eq.~\eqref{eq:CSD}) and the spherical-harmonic component of the overlap reduction function. 

To fully illustrate the power of our analysis, we apply our scheme to mock data containing (i) no GWB signal, (ii) a time-independent dipole GWB signal, and (iii) a GWB formed by Galactic plane binaries. 
We show that our analysis is capable of extracting the angular structures of all of these signals, despite each type corresponding to different levels of anisotropy. 
In particular, we show that, in the strong signal-to-noise ratio limit, our analysis can recover an accurate sky map that is almost identical to the simulated Galactic plane signal without bias. 
Through our Bayes factor calculations, we show that the data can be used to infer the suitable angular length scale of anisotropies that should be included in the analysis. 

Our marginalization scheme has several advantages compared to other existing search methods of anisotropic GWBs. 
First, the analytical formulae derived in this work allow us to reconstruct the intensity map of a GWB extremely accurately and rapidly, and compute the Bayes factors efficiently, completely bypassing the numerical sampling of an extremely high-dimensional posterior, which creates severe computational challenges.
Second, since our analysis does not require prior knowledge about the relationship between various spherical-harmonic components, our work represents a major step toward a model-independent search for anisotropic GWBs, which is crucial for understanding the properties of their sources. 

The remainder of this paper presents the details of the calculations summarized above, and it is organized as follows. 
Section~\ref{sec:Properties} lays the foundation of our analysis by first reviewing the basic properties of GWBs. 
Section~\ref{sec:Method} explains the method we develop and defines different probability distribution functions and hypothesis ranking for the Bayesian search of GWBs. 
Section~\ref{sec:marginalization} presents the details of the analytic marginalization and of the evaluation of the Bayes factor. 
Section~\ref{sec:Mock_data} applies the marginalization to mock data. 
Section~\ref{sec:Conclusion} concludes and points to future research. 
Throughout this paper, we adopt the following conventions:  bold lowercase characters represent a vector; bold uppercase characters represent a square matrix, and their corresponding italic unbolded characters with subscript(s) represent the elements of this matrix. 
Complex conjungation of a number is denoted by an asterisk. 
For example, $a_i$ is the $i$th element of the vector $\textbf{a}$, $A_{ij}$ is the $(i,j)$-th element of the square matrix $\textbf{A}$ and $a^{*}$ is the complex conjugate of $a$. 
Following \cite{LIGO_SGWB_01, LIGO_SGWB_02, KAGRA:2021kbb, KAGRA:2021mth}, we take the value of the Hubble constant to be $H_0 = 67.9 ~ \rm km s^{-1} Mpc^{-1}$, which is the \textit{Planck} 2015 value \cite{Planck:2015fie}, although our conclusions will not depend on this choice. 
While the analysis presented in this paper can be performed in any coordinate system, in this work we define the Sky position $\hat{\Omega}$ and analyze the anisotropy in celestial coordinates (in right accession and declination). 

\section{Properties of anisotropic gravitational-wave background}
\label{sec:Properties}

In this section, we will briefly review the properties and Bayesian analysis of an anisotropic GWB. 
Only GW properties that are strictly relevant to our work will be reviewed. 
We refer the reader to, for example, \cite{Thrane:2009fp, Romano:2016dpx, Romero:2021kby} for a more detailed and exhaustive review of anisotropic GWB. 

In general, metric perturbations at a given spacetime position can be written as a sum of contributions coming from all directions in the sky through a plane-wave expansion \cite{Allen:1996gp, Allen:1997ad, Romano:2016dpx}, 
\begin{equation}\label{eq:metric_perturbation}
\begin{split}
& h_{i j}(t, \mathbf{x})\\
& =\sum_{A=+, x} \int_{-\infty}^{\infty} d f \int d^{2} \hat{\Omega} ~ \tilde{h}_{A}(f, \hat{\Omega}) e_{i j}^{A}(\hat{\Omega}) e^{-2 \pi i f(t-\hat{\Omega} \cdot \mathbf{x})}, 
\end{split}
\end{equation}
where $A = +$ and $\times $ stand for the GW polarization, $\hat{\Omega}$ is a unit vector pointing in a sky direction, $e_{i j}^{A}(\hat{\Omega})$ are the GW polarization tensors, and the overhead tilde stands for the Fourier transform. 
Without loss of generality, the expectation value \footnote{To be more specific, if one assumes ergodicity, the expectation value is equivalent to the ensemble average, which is also the spatial average or the temporal average, see, e.g. \cite{Renzini:2022alw, Romano:2016dpx} for a more detailed review.} of $h_{ij}$ produced by a GWB is usually assumed to be zero \cite{maggiore2008gravitational}, 
\begin{equation}
\braket{h_{ij}(t, \textbf{x})} = 0. 
\label{eq:exp-value-hij}
\end{equation}
However, the quadratic expectation value of $\tilde{h}_{\rm A}$ is not zero \cite{Romano:2016dpx}, 
\begin{equation}
\begin{split}
& \left\langle\tilde{h}_{A}(f, \hat{\Omega}) \tilde{h}^{*}_{A^{\prime}} (f^{\prime}, \hat{\Omega}^{\prime})\right\rangle = \frac{1}{4}  \delta (f-f^{\prime}) \delta^{2} (\hat{\Omega}, \hat{\Omega}^{\prime}) \delta_{A A^{\prime}} \mathcal{H}(f, \hat{\Omega}), 
\end{split}
\end{equation}
where $\delta(\cdot)$ and $\delta^2(\cdot)$ are one- and two-dimensional Dirac delta functions respectively, $\delta_{A A'}$ is a Kronecker delta, and $\mathcal{H}(f, \hat{\Omega})$ is a function of frequency and the sky position  $\hat{\Omega}$ that is related to the one-sided strain power spectral density (PSD) of the GWB via \cite{Allen:1996gp, Thrane:2009fp, Romano:2016dpx}, 
\begin{equation}
S_h(f) = \frac{1}{4 \pi}\int_{S^2} d \hat{\Omega} \; \mathcal{H}(f, \hat{\Omega}). 
\end{equation}
In other words, $\mathcal{H}(f, \hat{\Omega})$ characterizes the angular distribution of the GWB power in different sky directions. 
The one-sided PSD is related to the dimensionless energy density (also known as the ``spectrum") of the GWB via
\begin{equation}
\Omega_{\mathrm{GW}}(f) \equiv \frac{f}{\rho_{\mathrm{c}}} \frac{d \rho_{\mathrm{GW}}}{d f}=\frac{8 \pi^3 }{3 H_0^2} f^3 S_h(f), 
\end{equation}
where $d \rho_{\mathrm{GW}}$ is the energy density of GWs of frequencies between $f$ and $f+df$, $\rho_c$ is the cosmological critical energy density ($\rho_c = 3 H_0^2 / 8 \pi G$) \footnote{Note that the normalization convention of $S_h(f)$ is different from that in some of the literature, like \cite{LIGOScientific:2011kvu, LIGOScientific:2016nwa, LIGOScientific:2019gaw, KAGRA:2021mth, Romano:2016dpx, Tsukada:2022nsu}. 
Here, we include a factor of $(4 \pi)^{-1}$ so that for the monopole part of GWB, we have $S_h(f) = \mathcal{H}_{00}(f)$, where $\mathcal{H}_{00}(f)$ is the monopole part of $\mathcal{H}(f; \hat{\Omega})$.}.
Closely related to $\Omega_{\rm GW}(f)$ is the power of the GWB per unit frequency per unit solid angle \cite{Romano:2016dpx, LIGOScientific:2016nwa, LIGOScientific:2019gaw}, 
\begin{equation}
\mathcal{F}(f, \Theta)=\frac{c^3 \pi}{4 G} f^2 \mathcal{H}(f, \hat{\Omega}),  
\end{equation}
where $c$ is the speed of light and $G$ is Newton's gravitational constant. 
As one would expect then, $\mathcal{F}(f, \Theta)$ has units of $\rm W~Hz^{-1}~sr^{-1}$. 

In general, the spectral content and the anisotropy of a GWB are correlated. 
However, the correlation may be difficult to measure with existing groud-based detectors, like advanced LIGO, advanced Virgo and KAGRA \cite{LIGOScientific:2011kvu, LIGOScientific:2019gaw, KAGRA:2021mth}. 
Hence, following past search on anisotropic GWBs \cite{LIGOScientific:2011kvu, LIGOScientific:2016nwa, LIGOScientific:2019gaw, KAGRA:2021mth}, we assume \cite{Romano:2016dpx, Thrane:2009fp, Tsukada:2022nsu, Allen:1996gp}
\begin{equation}
\mathcal{H}(f, \hat{\Omega}) = H(f) \mathcal{P}(\hat{\Omega}), 
\end{equation}
where $H(f)$ represents the spectral shape of the GWB and $\mathcal{P}(\hat{\Omega})$ encapsulates the strength and angular distribution of the intensity of the GWB, a function of the sky position. 
As in any other GWB search, we need to specify the spectral shape of the GWB, $H(f)$, that we are trying to detect. 
Within the sensitivity band of ground-based detectors, the energy spectrum of many GWBs can be well approximated by a power law in frequency \cite{LIGO_SGWB_01, LIGO_SGWB_02, LIGO_SGWB_03, LIGO_SGWB_04}, which means we can choose $H(f)$ to also have a power law structure, namely 
\begin{equation} \label{eq:H_alpha}
H(f) = H_{\alpha} (f) = \left( \frac{f}{f_{\rm ref}}\right)^{\alpha-3}, 
\end{equation}
where $f_{\rm ref}$ is a reference frequency and $\alpha$ is the tilt index.
Following the existing search of GWB from the actual data, we will fix $\alpha$ and infer $\mathcal{P}(\hat{\Omega})$. 
Throughout this paper, we also follow the existing LIGO/Virgo search of a GWB and choose $f_{\rm ref} = 25 \rm Hz$, $\alpha = 0, 2/3$ or 3\cite{LIGO_SGWB_01, LIGO_SGWB_02, LIGO_SGWB_03, LIGO_SGWB_04}. The choice of $f_{\rm ref}$ does not affect the rest of the analysis at all because it just provides the overall normalization for $\Omega_{\rm GW}$. 

To extract the angular structure of the GWB from data, we perform a spherical-harmonic decomposition to express $\mathcal{P}(\hat{\Omega})$ as a linear combination of (scalar) spherical harmonics $Y_{\ell m}(\hat{\Omega})$, 
\begin{equation}\label{eq:spherical harmonic decomposition}
\mathcal{P}(\hat{\Omega}) = \sum_{\ell = 0}^{\ell_{\rm max}} \sum_{m = - \ell}^{+\ell} \mathcal{P}_{\ell m} Y_{\ell m}(\hat{\Omega}), 
\end{equation}
where $\mathcal{P}_{\ell m}$ are referred to as \textit{spherical-harmonic components} of the spectrum of the GWB, in units of $\rm strain^2 \; Hz^{-1} \, rad^{-1}$. 
In principle, this sum must include an infinite number of $\ell$ terms, but in practice, one must truncate the sum at some $\ell = \ell_{\rm max}$.
The value of $\ell_{\rm max}$ will be specified in subsequent calculations. 
From Eq.~\eqref{eq:spherical harmonic decomposition}, we notice two things. 
First, upon sky averaging (i.e.~integration over sky angle), all terms vanish except $\mathcal{P}_{00}$. 
This implies that 
\begin{equation}
\mathcal{P}_{00} = \frac{S_h(f_{\rm ref})}{\sqrt{4 \pi}} = \frac{3 H_0^2}{2 \pi^2 f_{\rm ref}^{3}} \frac{\Omega_{\rm GW} (f_{\rm ref})}{\sqrt{4 \pi}}. 
\end{equation}
Second, the real nature of $ \mathcal{P}(\hat{\Omega}) $ and $Y_{\ell 0}$ and the complex conjugation property of $Y_{\ell, m}$ imply that
\begin{equation}
\begin{split}
Y_{\ell 0} \in \mathbb{R} & \Rightarrow \mathcal{P}_{\ell 0} \in \mathbb{R}, \\
Y_{\ell, -m} =(-1)^m Y^{*}_{\ell m} & \Rightarrow \mathcal{P}_{\ell, -m} =(-1)^m \mathcal{P}^{*}_{\ell m}.  
\end{split}
\end{equation}
These requirements imply that, in order to specify the angular distribution of a GWB, we only need $(\ell_{\max}+1)^2$ real numbers, 
\begin{equation}
\begin{split}
& \mathcal{P}_{0 0}, \mathcal{P}_{1 0}, ...., \mathcal{P}_{\ell 0}, \\
& \mathcal{P}^{\rm Re}_{1 1}, \mathcal{P}^{\rm Re}_{2 1}, ...., \mathcal{P}^{\rm Re}_{\ell m}, \\
& \mathcal{P}^{\rm Im}_{1 1}, \mathcal{P}^{\rm Im}_{2 1}, ...., \mathcal{P}^{\rm Im}_{\ell m}, 
\end{split}
\end{equation}
where $\mathcal{P}^{\rm Re}_{\ell m}$ and $\mathcal{P}^{\rm Im}_{\ell m}$ are, respectively, the real and imaginary parts of $\mathcal{P}_{\ell m}$.
For the sake of clarity, we introduce a $(\ell_{\max}+1)^2$-vector to denote these numbers, 
\begin{equation}
\begin{split}
\textbf{w} = & (\mathcal{P}_{0 0}, \mathcal{P}_{1 0}, ...., \mathcal{P}_{\ell 0}, \\
& \mathcal{P}^{\rm Re}_{1 1}, \mathcal{P}^{\rm Re}_{2 1}, ...., \mathcal{P}^{\rm Re}_{\ell m}, \\
& \mathcal{P}^{\rm Im}_{1 1}, \mathcal{P}^{\rm Im}_{2 1}, ...., \mathcal{P}^{\rm Im}_{\ell m})^{\rm T}. 
\end{split}
\end{equation}

\section{Basics of Bayesian search for gravitational-wave background}
\label{sec:Method}

Searching for an anisotropic GWB amounts to determining the spherical-harmonic components of the intensity map from the data. 
In the presence of overwhelming noise, the spherical-harmonic components, just as the estimation of the parameters of other GW signals, should be determined by Bayesian inference. 
This section is devoted to reviewing the basics of a Bayesian search for an anisotropic GWB. 
Before we explain the Bayesian strategy we propose, we will first state the assumptions and simplifications we make for the calculations, and then we will justify them.
Then, we will define the likelihood, prior and posterior for the Bayesian search. 
In terms of the spherical-harmonic components defined in the last section, we will explicitly write down the likelihood function as a Gaussian of the real and imaginary parts of the spherical-harmonic components of the GWB. 
Finally, we also define the Bayes factor, which competes the hypothesis that an anisotropic GWB is detected against that hypothesis that the data consists of only noises. 

\subsection{Approximations and simplifications}

To construct various probability distribution functions that will be used in our Bayesian GWB analysis, we make the following assumptions and approximations: 
\begin{enumerate}[label=A.\arabic*,start=1]
    \item The GWB signal is weak, in the sense that the auto-correlated power and the cross-correlated power of the GWB signal are much smaller than that of the detector noise. 
    This assumption is justified because the current observational constraints on the strength of GWB indicates that, if a GWB is present at all, it must be weak \cite{KAGRA:2021kbb, KAGRA:2021mth}. 
    \item The instrumental noise is stationary and Gaussian. 
    In practice, this assumption is not always realistic as individual GW signals or non-Gaussian noise transients, known as glitches, will occasionally be present. 
    Nonetheless, data segments containing non-Gaussian transients will be removed by applying data cuts \cite{LIGO_data_cut_04}. 
    Hence, in our analysis, we can assume that the detector response is Gaussian. 
    \item The noise across different detectors is not correlated. 
    This assumption is realistic because the existing ground-based detectors are spatially well-separated. 
    \item We assume that different segments of the data are independent (uncorrelated in time and frequency) to simplify the statistical calculations. 
    This assumption implies that the likelihood function can be written as a product of the likelihood over different time and frequency segments. 
    In practice, the segments are correlated for two reasons. 
    First, the serial dependence in the entire segment of the time-domain data introduces auto-correlations over time and frequency. 
    Second, when we transform time-domain data segments into the frequency-domain, the data segments are windowed. 
    To make full use of the windowed data, the time-domain data segements need to overlap, which introduces correlations between them. 
    In practice, to address these correlations, Eq.~\eqref{eq:mu_i} (one of our key results) should first be applied to each individual segment (with appropriate windowing factors multiplied), and then optimally combined \cite{Lazzarini:2004hk, Overlapping_windows, bias_factor}. 
    Reference \cite{Dipole_01}, however, has shown that, after taking all these correlations into account, the optimally-combined results from all data segments agree well with predictions obtained from a likelihood that assumes the data segments are independent of each other. Such consistency justifies the simplifications used in this paper. 
\end{enumerate}

\subsection{Likelihood and posterior in Bayesian GWB analysis}
\label{sec:likelihood}

When a GWB is present, it induces responses on GW detectors that force the latter to measure strain data consisting of two parts: 
\begin{equation}\label{eq:TD_strain}
\tilde{s}_I(f, t) = \tilde{n}_I (f, t) + \tilde{h}_I (f, t), 
\end{equation}
where $I$ labels the detector and $\tilde{s}_I(f, t)$ is the finite- or short-time Fourier transform of the time-domain data in detector $I$, 
$s_I (t)$, within a time interval $[t-\tau/2, t+\tau/2]$, 
\begin{equation}
\tilde{s}_I(f, t) = \int_{t-\tau/2}^{t+\tau/2} dt' w(t') s_I(t') e^{-2i\pi ft'}, 
\end{equation}
with $w(t)$ a windowing function. 
Similarly, $\tilde{n}_I (f, t)$ and $\tilde{h}_I (f, t)$ are the short-time Fourier transform of the instrumental noise , $n_I(t)$, and of the GWB-induced response, $h_I(t)$, of detector $I$, respectively. 
The time-domain GWB-induced response on detector $I$ is related to the metric perturbations of the GWB [Eq.~\eqref{eq:metric_perturbation}] via 
\begin{equation}
h_I (t) = D_I^{ij} h_{ij}(t, \textbf{x}_I), 
\end{equation}
where $D_I^{ij}$ is a tensor (known as the detector response tensor) that encapsulates the geometry and orientations geometry of detector $I$, and $\textbf{x}_I$ is the position vector of detector $I$. 

The expectation value of the GWB-induced response satisfies 
\begin{equation}
\braket{\tilde{h}_I (f, t)} = 0\,, 
\end{equation}
which descends directly from Eq.~\eqref{eq:exp-value-hij}. 
As $\tilde{h}_I(f, t)$ is random and has zero mean, the GWB-induced response just looks like noise within individual detectors.
However, the responses induced on two GW detectors, say $I$ and $J$, should be proportional to each other (in the time domain), which means that they should be correlated across among detectors \cite{Thrane:2009fp}, 
\begin{equation}
\braket{\tilde{h}_I (f, t) \tilde{h}^{*}_J (f, t)} = \frac{\tau}{2} H_{\alpha} (f) \sum_{\ell m} \gamma^{(IJ)}_{\ell m}(f, t) \mathcal{P}_{\ell m}, 
\end{equation}
where $\tau$ is the time length of the data segment analyzed, and $\gamma_{\ell m}^{(IJ)} (f, t)$ is the spherical-harmonic components of the overlap reduction function (ORF)
of detectors $I$ and $J$, defined by \cite{Romano:2016dpx}
\begin{equation}\label{eq:ORF_SHCs}
\begin{split}
& \gamma^{(IJ)}(f, t, \hat{\Omega}) = \frac{1}{2} \sum_{A} R^{(I)}_{A} (f, t, \hat{\Omega}) \left[ R^{(J)}_{A} (f, t, \hat{\Omega}) \right]^{*}, \\
& \gamma^{(IJ)}_{\ell m}(f, t) = \int d^{2} \hat{\Omega} \; \gamma^{(IJ)}(f, t, \hat{\Omega}) \; Y_{\ell m}(\hat{\Omega}), 
\end{split}
\end{equation}
where $A = +, \times$ stands for the GW polarization, and $R^{(I)}_{A} (f, t, \hat{\Omega})$ is the polarization-basis response function of detector $I$. The latter depends on time because of Earth's rotation. 
As the definition suggests, $\gamma^{(IJ)}_{\ell m}(f, t)$ encapsulates information about the detectors' geometry, location, orientation and antenna pattern, and they should not be confused with the spherical-harmonic components of the spectrum of the GWB, $\mathcal{P}_{\ell m}$. 
Instrumental noise, on the other hand, has very different properties: 
if $I$ and $J$ are well separated, their instrumental noise should be uncorrelated, 
\begin{equation}\label{eq:noise_corr_diff_detectors}
\braket{\tilde{n}_I (f, t) \tilde{n}^{*}_J (f, t)} = 0. 
\end{equation}
By the same token 
\begin{equation}\label{eq:zero_corr}
\begin{split}
& \braket{\tilde{n}_I (f, t) \tilde{h}^{*}_J (f, t)} = \braket{\tilde{n}_J (f, t) \tilde{h}^{*}_I (f, t)} \\
& = \braket{\tilde{n}_I (f, t) \tilde{h}^{*}_I (f, t)} = \braket{\tilde{n}_J (f, t) \tilde{h}^{*}_J (f, t)} = 0. 
\end{split}
\end{equation}

These correlation properties suggest that a GWB can be searched for by cross-correlating the strain data measured by different detectors. 
To this end, we define the cross-spectral density, $C(f, t)$, between two detectors, $I$ and $J$, via 
\begin{equation}\label{eq:CSD}
C(f, t) \equiv \frac{2}{\tau} \tilde{s}_I(f, t) \tilde{s}_J^{*} (f, t).
\end{equation}
If a GWB is present, then the expectation value of $C(f, t)$ is \cite{Thrane:2009fp}, 
\begin{equation}
\begin{split}
\langle C(f, t)\rangle & = H_{\alpha} (f) \sum_{\ell m} \gamma_{\ell m}(f, t) \mathcal{P}_{\ell m}. 
\end{split}
\end{equation}
We can also derive the (approximate) variance of $C(f, t)$ for a weak GWB signal by considering the covariance matrix, 
\begin{equation}
\begin{split}
\text{Cov}(f, t; f', t') & = \braket{C(f, t) C^{*}(f', t')} - \braket{C(f, t)}\braket{C^{*}(f', t')} \\
& \approx \braket{C(f, t) C^{*}(f', t')}, 
\end{split}
\end{equation}
where we have dropped the term $\braket{C(f, t)}\braket{C^{*}(f', t')}$ because it corresponds to a second-order contribution in the weak-signal approximation (c.f. A.1.). 
Then, using Eqs.~\eqref{eq:CSD}, \eqref{eq:TD_strain}, \eqref{eq:noise_corr_diff_detectors} and \eqref{eq:zero_corr}, we have 
\begin{equation}\label{eq:covariance_weak_signal}
\begin{split}
& \text{Cov}(f, t; f', t')
= \delta_{ff'} \delta_{tt'} N_I(f, t) N_J(f, t), 
\end{split}
\end{equation}
where $N_{I, J}(f, t)$ is the one-sided PSD of the output of detectors $I$ and $J$. 
We remind the reader that the derivation is valid only if we assume A.1.
Reference \cite{Matas:2020roi} has verified that Eq.~\eqref{eq:covariance_weak_signal} gives an accurate estimate of the variance of the cross-correlation for the search of isotropic GWBs. 
Since anisotropic GWBs is an extension of an isotropic GWB search, we expect that Eq.~\eqref{eq:covariance_weak_signal} remains accurate in this case as well. 

The Bayesian search of an anisotropic GWB amounts to determining the posterior of $\textbf{w}$, given the cross-spectral density, of all data segments $\left\{ C\right\}$. 
According to Bayes' theorem, the posterior is related to the likelihood by
\begin{equation}\label{eq:Bayes_Thm}
p(\textbf{w}| \left\{ C\right\}, H) = \frac{p(\textbf{w}|H) p(\left\{ C\right\} |\textbf{w}, H)}{p(\left\{ C\right\}|H)}. 
\end{equation}
Here, $p(\textbf{w}| \left\{ C\right\}, H)$ is the posterior of $\textbf{w}$, given the cross-spectral density and the hypothesis $H$ (e.g.~that the measured data contain a GWB signal, which will be more precisely defined in Eq.~\eqref{eq:hypothese}). The quantity $p(\left\{ C\right\}|H)$ is the Bayesian evidence, which is a normalization constant of the posterior.  The quantity $p(\textbf{w}|H)$ is the prior of $\textbf{w}$, prescribed according to our hypothesis. The quantity
$p(\left\{ C\right\} |\textbf{w}, H)$ is the likelihood that we will measure $\left\{ C\right\}$, given that there is a GWB with spherical-harmonic components $\textbf{w}$. 
Using the weak-signal approximation, and the expectation value and the variance derived above, the likelihood $p(\left\{ C\right\} |\textbf{w}, H)$ can be modeled by \cite{Thrane:2009fp, Romano:2016dpx, Tsukada:2022nsu}
\begin{equation}\label{eq:likelihood}
\begin{split}
& p(\left\{ C\right\} |\textbf{w}, H) \\
& = \mathcal{N} \exp \left\{-\frac{1}{2} \sum_{f, t} \frac{\left|C(f, t)- H_{\alpha}(f) \gamma_{\mu} \mathcal{P}_{\mu}\right|^2}{N_I(f, t) N_J(f, t)}\right\}, 
\end{split}
\end{equation}
where $\sum_{f,t}$ stands for summation over frequency bins and the center times of the short-timed Fourier transform, $\mathcal{N} $ is a proportionality constant that does not depend on $\mathcal{P}_{\ell m}$, and $ \gamma_{\mu} \mathcal{P}_{\mu}$ is shorthand notation for 
\begin{equation}
\gamma_{\mu} \mathcal{P}_{\mu} = \sum_{\ell = 0}^{\ell_{\rm max}^{\rm (inf)}} \sum_{m = - \ell}^{+ \ell} \gamma_{\ell m}(f, t) \mathcal{P}_{\ell m}, 
\end{equation}
where $\mu = (\ell, m)$ labels the mode, 
and $\ell_{\rm max}^{\rm (inf)}$ is the maximum $\ell$ that we include in the inference analysis. 

When searching for an anisotropic GWB, Eq.~\eqref{eq:Bayes_Thm} represents a high-dimensional posterior probability distribution function, which is difficult to visualize.
Thus, it is very convenient to present the marginalized posterior of a particular spherical-harmonic component. 
To this end, one can marginalize the posterior (Eq.~\eqref{eq:Bayes_Thm}) over all components of $\textbf{w}$ that one is not interested in (at the moment) to obtain the marginalized posterior of, say, $w_i$, 
\begin{equation}
\begin{split}
& p(w_i| \left\{ C\right\}, H) = \prod_{j \neq i} \int d w_j p(\textbf{w}| \left\{ C\right\}, H). 
\end{split}  
\end{equation}
The lower and upper limit of the integral involved in the marginalization depends on $p(\textbf{w}|H)$, which will be prescribed in Sec.~\ref{sec:marginalization}. 

\subsection{The Bayes factor}

Other than constructing the marginalized posterior, Bayesian theory also provides a framework to compute the so-called ``Bayes factor.'' The latter is a measure that allows one to compare two hypotheses in light of the data within Bayesian inference. In the context of GWBs, the Bayes factor can be used to quantify whether an anisotropic GWB has been detected or not by comparing the following two hypotheses:
\begin{equation}\label{eq:hypothese}
\begin{split}
& H_{\ell_{\rm max}}: \text{the data $\left \{ C\right\}$ contain a GWB signal whose} \\
& \quad \quad \quad \text{$\mathcal{P}_{\ell_{\rm max} m} \neq 0$ for at least one $m \in $} \\
& \quad \quad \quad \text{$[ - \ell_{\rm max}, -\ell_{\rm max} + 1, ..., \ell_{\rm max} - 1, \ell_{\rm max} ]$ , and } \\
& H_{\rm null}: \text{the data $\left \{ C\right\}$ contain only noise}. 
\end{split}
\end{equation} 
In Bayesian inference, we can compare these two hypotheses by computing their odds ratio, namely, the ratio of their respective evidences given the data:
\begin{equation}
\mathcal{O}(\ell_{\rm max}) = \frac{p(H_{\ell_{\rm max}}|\{ C\})}{p(H_{\rm null}|\{ C\})} =  \frac{p(H_{\ell_{\rm max}})}{p(H_{\rm null})} \frac{p(\{ C\}|H_{\ell_{\rm max}})}{p(\{ C\}|H_{\rm null})}. 
\end{equation}
The term ${p(H)}/{p(H_{\rm null})}$ is known as the prior odds, and it represents our prior belief of one hypothesis over the other. The second term in the above equation is known as the Bayes factor, 
\begin{equation}
\mathcal{B}(\ell_{\rm max}) = \frac{p(\{ C\}|H_{\ell_{\rm max}})}{p(\{ C\}|H_{\rm null})}\,,
\end{equation}
which implies that the odds ratio is nothing but the product of the prior odds with the Bayes factor. 

One can think of the Bayes factor as the odds ratio between two hypotheses under the assumption of equal prior belief between them. 
As we have no information about whether we have detected a GWB before we analyze the data, we naturally assume the two hypotheses are equally likely.
Thus, 
\begin{equation}
p(H_{\ell_{\rm max}}) = p(H_{\rm null}) \Rightarrow \mathcal{O}(\ell_{\rm max}) = \mathcal{B}(\ell_{\rm max}). 
\end{equation}
If $\mathcal{B}(\ell_{\rm max}) > 1$, then hypothesis $H_{\ell_{\rm max}}$ is favored over hypothesis $H_{\rm null}$, which implies it is more likely that we have detected a GWB than not; the opposite is true, of course, if $\mathcal{B}(\ell_{\rm max}) \leq 1$. For convincing evidence that we have indeed detected a GWB, one typically requires that $\mathcal{B}(\ell_{\rm max}) \gg 1$, where precisely how much larger than unity this requirement must depend on the statistician's definition of ``convincing'' \cite{Romano:2016dpx}. 

\section{Analytic Marginalization of the Posterior and Bayes Factor}
\label{sec:marginalization}

As pointed out in the last section, the posterior of the spherical-harmonic components is a probability distribution function of high dimension. 
In principle, one can numerically sample the posterior using nested sampling or Markov Chain Monte Carlo techniques. 
But given the high dimensionality of the distribution, both sampling approaches will take an extremely long time to complete in the GWB case. 
In this section, we will show that, if a wide-enough uniform prior is prescribed, the marginalized posterior and Bayes factor for the search for an anisotropic GWB can be \textit{analytically} evaluated as a high-dimensional Gaussian integral, with the former also a Gaussian function. 

\subsection{Marginalized posterior}

Let us begin by explicitly writing down the exponent of the likelihood as a quadratic form of $\textbf{w}$. 
To start, we rewrite the likelihood as 
\begin{equation}
p(\left\{ C\right\} |\textbf{w}, H_{\ell^{\rm (inf)}_{\rm max}}) \propto \exp \left\{-\frac{1}{2} \sum_{f, t} \frac{R(f, t) R^{*}(f, t)}{N_I(f, t) N_J(f, t)} \right\}
\end{equation}
where $R(f, t)$ stands for the residual
\begin{equation}
R(f, t) \equiv C(f, t)- H(f) \sum_{\ell=0}^{\ell^{\rm (inf)}_{\rm max}} \sum_{m=-\ell}^{\ell} \gamma_{\ell m}(f, t) \mathcal{P}_{\ell m}. 
\end{equation}
\begin{widetext}
Explicitly writing out the summation over $\ell$ and $m$, we have 
\begin{equation}
\begin{split}
\sum_{\ell=0}^{\ell_{\rm max}^{\rm (inf)}} \sum_{m=-\ell}^{+\ell} \gamma_{\ell m}(f, t) \mathcal{P}_{\ell m} & = \sum_{\ell=0}^{\ell_{\rm max}^{\rm (inf)}} \gamma_{\ell 0}(f) \mathcal{P}_{\ell 0} + \sum_{\ell=0}^{\ell_{\rm max}^{\rm (inf)}} \sum_{m=1}^{+\ell} [ \gamma_{\ell m}(f, t) \mathcal{P}_{\ell m} + \gamma_{\ell, -m}(f, t) \mathcal{P}_{\ell, -m} ] \\
& = \sum_{\ell=0}^{\ell_{\rm max}^{\rm (inf)}} \gamma_{\ell 0}(f) \mathcal{P}_{\ell 0} + \sum_{\ell=0}^{\ell_{\rm max}^{\rm (inf)}} \sum_{m=1}^{+\ell} [ \gamma_{\ell m}(f, t) \mathcal{P}_{\ell m} + (-1)^{\ell}\gamma^{*}_{\ell m}(f, t) \mathcal{P}^{*}_{\ell m} ], 
\end{split}
\end{equation}
where in the last line, we have used Eqs.~(B1) and (B4) of \cite{Thrane:2009fp}, namely
\begin{equation}
\begin{split}
\gamma_{\ell m}^{*} (f, t) & =(-1)^{\ell+m} \gamma_{\ell,-m}(f, t), \\
\mathcal{P}_{\ell m}^{*} & =(-1)^{m} \mathcal{P}_{\ell,-m}. 
\end{split}
\end{equation}

We further decompose $\gamma_{\ell m} \mathcal{P}_{\ell m}$ into its real and imaginary parts, 
\begin{equation}
\begin{split}
& \Re \left[ \gamma_{\ell m} \mathcal{P}_{\ell m} \right] = \sum_{\ell=0}^{\ell_{\rm max}^{\rm (inf)}} \gamma^{\rm Re}_{\ell 0}(f) \mathcal{P}_{\ell 0} + \sum_{\ell=0}^{\ell_{\rm max}^{\rm (inf)}} \sum_{m=1}^{+\ell} \left[1+(-1)^{\ell}\right] [ \gamma^{\rm Re}_{\ell m}(f, t) \mathcal{P}^{\rm Re}_{\ell m} - \gamma^{\rm Im}_{\ell m}(f, t) \mathcal{P}^{\rm Im}_{\ell m} ], \\
& \Im \left[ \gamma_{\ell m} \mathcal{P}_{\ell m} \right] = \sum_{\ell=0}^{\ell_{\rm max}^{\rm (inf)}} \gamma^{\rm Im}_{\ell 0}(f) \mathcal{P}_{\ell 0} + \sum_{\ell=0}^{\ell_{\rm max}^{\rm (inf)}} \sum_{m=1}^{+\ell} \left[1-(-1)^{\ell}\right] [ \gamma^{\rm Im}_{\ell m}(f, t) \mathcal{P}^{\rm Re}_{\ell m} + \gamma^{\rm Re}_{\ell m}(f, t) \mathcal{P}^{\rm Im}_{\ell m} ].
\end{split}
\end{equation}
These expressions can be more compactly expressed if we define two $(\ell^{\rm (inf)}_{\rm max}+1)^2$ vectors, $\textbf{u}(f, t)$ and $\textbf{v}(f,t)$ such that
\begin{equation}
\begin{split}
\Re \left[R(f, t)\right] &= C^{\rm Re} (f, t) - \textbf{u}^{\rm T} (f, t) \cdot \textbf{w}, \\
\Im \left[R(f, t)\right] &= C^{\rm Im} (f, t) - \textbf{v}^{\rm T} (f, t) \cdot \textbf{w}, 
\end{split}
\end{equation}
where 
\begin{equation}
\begin{split}
\textbf{u} (f, t) \equiv & H_{\alpha}(f) (\gamma^{\rm Re}_{0 0}, \gamma^{\rm Re}_{1 0}, ...., \gamma^{\rm Re}_{\ell 0}, \left[1+(-1)^{1}\right] \gamma^{\rm Re}_{1 1}, \\
& \left[1+(-1)^{2}\right] \gamma^{\rm Re}_{2 1}, ...., \left[1+(-1)^{\ell}\right] \gamma^{\rm Re}_{\ell m}, -\left[1+(-1)^{1}\right]\gamma^{\rm Im}_{1 1}, -\left[1+(-1)^{2}\right]\gamma^{\rm Im}_{2 1}, ...., -\left[1+(-1)^{\ell}\right]\gamma^{\rm Im}_{\ell m})^{\rm T}, \\
\textbf{v} (f, t) \equiv & H_{\alpha}(f) (\gamma^{\rm Im}_{0 0}, \gamma^{\rm Im}_{1 0}, ...., \gamma^{\rm Im}_{\ell 0}, \left[1-(-1)^1\right] \gamma^{\rm Im}_{1 1}, \\
& \left[1-(-1)^{2}\right] \gamma^{\rm Im}_{2 1}, ...., \left[1-(-1)^{\ell}\right] \gamma^{\rm Im}_{\ell m}, \left[1-(-1)^{1}\right] \gamma^{\rm Re}_{1 1}, \left[1-(-1)^{2}\right] \gamma^{\rm Re}_{2 1}, ...., \left[1-(-1)^{\ell}\right] \gamma^{\rm Re}_{\ell m})^{\rm T}. 
\end{split}
\end{equation}
\end{widetext}
Note that $\textbf{u}(f, t)$ and $\textbf{v}(f,t)$ depend only on the ORF of the detectors, but each element of $\textbf{u}(f, t)$ and $\textbf{v}(f,t)$ is a function of $f$ and $t$ because they inherit the frequency and time dependence from $H(f)$ and $\gamma_{\ell m}(f, t)$.

With $\textbf{u}$ and $\textbf{v}$ defined, the square of the modulus of $ R(f, t) $ can be computed as a quadratic function of $\textbf{w}$
\begin{equation}
\begin{split}
& R(f, t) R^{*}(f, t) \\
& = C(f, t) C^{*}(f, t) - 2 \textbf{g}^{\rm T} (f, t) \cdot \textbf{w} + \textbf{w}^{\rm T} \cdot \textbf{K} (f, t) \cdot \textbf{w}, 
\end{split}
\end{equation}
where $\textbf{g}(f, t)$ is a $(\ell^{\rm (inf)}_{\rm max}+1)^2$-vector, 
\begin{equation}
\textbf{g}(f, t) \equiv C^{\rm Re}(f, t) \textbf{u}(f, t) + C^{\rm Im}(f, t) \textbf{v}(f, t), 
\end{equation}
and $\textbf{K}(f, t)$ is a symmetric-square matrix of order of $(\ell^{\rm (inf)}_{\rm max}+1)^2$, whose elements are given by 
\begin{equation}
K_{ij} (f, t) \equiv u_i u_j + v_i v_j. 
\end{equation}
Note that while $\textbf{g}(f, t)$ depends on the data via the real and imaginary parts of $C(f, t)$, $K_{ij}$ depends solely on the detectors' geometry via the dependence on the ORF. 
Similarly, we can also write the exponent of the likelihood as a quadratic function of $\textbf{w}$, 
\begin{equation}
\begin{split}
& \sum_{f, t} \frac{R(f, t) R^{*}(f, t)}{N_I(f, t) N_J(f, t)} \\
& = \sum_{f, t} \frac{|C(f, t)|^2}{N_I(f, t) N_J(f, t)} - 2 \textbf{j}^{\rm T} \cdot \textbf{w} + \textbf{w}^{\rm T} \cdot \textbf{Q} \cdot \textbf{w}, 
\end{split}
\end{equation}
where $\textbf{j}$ is a $[\ell_{\rm max}^{\rm (inf)}+1]^2$-vector and $\textbf{Q}$ is another symmetric-square matrix of order of $[\ell_{\rm max}^{\rm (inf)}+1]^2 $. 
Explicitly, their elements are
\begin{equation}\label{eq:j_and_Q}
\begin{split}
j_i & \equiv \sum_{f, t} \frac{g_i (f, t)}{N_I (f, t) N_J(f, t)}, \\
Q_{i j} & \equiv \sum_{f, t} \frac{K_{ij} (f, t)}{N_I (f, t) N_J(f, t)}. 
\end{split}
\end{equation}
Recall that $\textbf{g}$ depends on the data, and so does $\textbf{j}$. 
Eventhough $\textbf{K}$ does not depend on data,  $\textbf{Q}$ does because it contains the PSD of the data. 
Unlike $\textbf{g}$ and $\textbf{K}$, $\textbf{j}$ and $\textbf{Q}$ are constant.
In terms of $\textbf{j}, \textbf{w}$ and $\textbf{Q}$, the likelihood and posterior are, respectively, given by 
\begin{equation}
\begin{split}
& p(\left\{ C\right\} |\textbf{w}, H_{\ell^{\rm (inf)}_{\rm max}}) = \bar{\mathcal{N}} \exp \left( \textbf{j}^{\rm T} \cdot \textbf{w} - \frac{1}{2} \textbf{w}^{\rm T} \cdot \textbf{Q} \cdot \textbf{w} \right), \\
& p(\textbf{w}| \left\{ C\right\}, H_{\ell^{\rm (inf)}_{\rm max}}) = \bar{\mathcal{N}} \frac{p(\textbf{w}|H_{\ell^{\rm (inf)}_{\rm max}})}{p(\left\{ C\right\}|H_{\ell^{\rm (inf)}_{\rm max}})} \\
& \quad \quad \quad \quad \quad \quad \quad \quad \quad \quad \times \exp \left( \textbf{j}^{\rm T} \cdot \textbf{w} - \frac{1}{2} \textbf{w}^{\rm T} \cdot \textbf{Q} \cdot \textbf{w} \right), 
\end{split} 
\end{equation}
where 
\begin{equation}
\bar{\mathcal{N}} = \mathcal{N} \exp \left( - \frac{1}{2}\sum_{f, t} \frac{|C(f, t)|^2}{N_I(f, t) N_J(f, t)} \right). 
\end{equation}
At this point, let us summerize and remind the reader that $\textbf{g}, \textbf{j} $ and $\textbf{Q}$ depend on the data, whereas $\textbf{u}, \textbf{v} $ and $\textbf{K}$ depend only on the geometry of the detectors via the ORF. 
Thus, $\textbf{u}, \textbf{v} $ and $\textbf{K}$ can be pre-computed and stored for given detectors to speed up the analysis. 

We are now ready to marginalize the posterior. 
If we are particularly interested in knowing the posterior of $w_i$, then the argument of the exponential of the posterior can be written as
\begin{equation}\label{eq:exponent}
\begin{split}
& \textbf{j}^{\rm T} \cdot \textbf{w} - \frac{1}{2} \textbf{w}^{\rm T} \cdot \textbf{Q} \cdot \textbf{w} \\
= &  j_i w_i - \frac{1}{2} Q_{i i} w_i^2 \\
& + \sum_{k \neq i} \left[j_k - \frac{1}{2} w_i (Q_{k i} + Q_{i k})\right]w_k - \frac{1}{2} \sum_{k \neq i} \sum_{l \neq i} w_k Q_{k l} w_l, 
\end{split}
\end{equation}
where the index $i$ in the first two terms, $j_i w_i$ and $Q_{i i} w_i^2$, does not imply summation. 
To facilitate subsequent calculations, we define the following $[\ell_{\rm max}^{\rm (inf)}+1]^2 -1$ vectors and square matrices of order $[[\ell_{\rm max}^{\rm (inf)}+1]^2 -1]$: 
\begin{align*}\label{eq:quantaties}
\tilde{\textbf{w}}^{(i)} = & \text{the vector $\textbf{w}$ with the $i$th element removed}, \\
\textbf{b}^{(i)} = & \text{the vector $\textbf{j}$ with the $i$th element removed}, \\
\textbf{a}^{(i)} = & \text{a vector whose $k$-th element is} \\
a_k = & Q_{i k} ~~ (\text{for $k \neq i$ given $i$}) \\
\textbf{n}^{(i)} = & \textbf{b}^{(i)} - w_i \textbf{a}^{(i)}, \\
\tilde{\textbf{Q}}^{(i)} = & \text{the matrix $\textbf{Q}$ with the $i$th row and  the $i$th} \\
& \text{column removed}, \\
\textbf{M}^{(i)} = & \text{the inverse of $\tilde{\textbf{Q}}^{(i)}$}. 
\end{align*}
Note that, except $\textbf{w}$, all these vectors and matrices depend on the data. The vectors
$\textbf{b}$ and $\textbf{n}$ depend on the data because $\textbf{j}$ depends on the data (c.f. Eq.~\eqref{eq:j_and_Q}). 
The vector $\textbf{a}$ and the matrices $\tilde{\textbf{Q}}^{(i)}$ and $\textbf{M}^{(i)} $ all depend on the data because $\textbf{Q}$ depends on the PSD of the data.  
Note also that, since $\tilde{\textbf{Q}}^{(i)}$ is symmetric, so is $\textbf{M}^{(i)}$. 
The argument of the exponential of the posterior can then be more compactly written as 
\begin{equation}
\begin{split}
& \textbf{j}^{\rm T} \cdot \textbf{w} - \frac{1}{2} \textbf{w}^{\rm T} \cdot \textbf{Q} \cdot \textbf{w} \\
= &  j_i w_i - \frac{1}{2} Q_{i i} w_i^2 + {\textbf{n}^{(i)}}{}^{\rm T} \cdot \tilde{\textbf{w}}^{(i)} - \frac{1}{2} {\tilde{\textbf{w}}^{(i)}}{}^{\rm T} \cdot \tilde{\textbf{Q}}^{(i)} \cdot \tilde{\textbf{w}}^{(i)}\,,
\end{split}
\end{equation}
where we recall that $\textbf{n}^{(i)}$ depends on both the index $i$ and $w_i$, and the repeated subscript $i$ does not imply summation. 

The posterior can be analytically marginalized if we choose a prior for $\textbf{w}$ with the following properties:
\begin{enumerate}
    \item The prior is factorized as a product of the prior of individual $w_i$, 
    \begin{equation}\label{eq:prior_w_i}
    p(\textbf{w}|H_{\ell^{\rm (inf)}_{\rm max}}) = \prod_{i=1}^{[\ell_{\rm max}^{\rm (inf)}+1]^2} p_i (w_i|H_{\ell^{\rm (inf)}_{\rm max}}), 
    \end{equation}
    where $p_i (w_i|H_{\ell^{\rm (inf)}_{\rm max}})$ is the prior of $w_i$. 
    By choosing a factorized prior for $\textbf{w}$, we are assuming that different $w_i$ are independent of each other. 
    \item Each $p_i (w_i|H_{\ell^{\rm (inf)}_{\rm max}})$ is uniform for $w_i \in [-\Delta^{(i)}, \Delta^{(i)}]$, where $\Delta^{(i)} > 0 $ is the width of the prior of $w_i$. 
    \item When $w_i = \pm \Delta^{(i)} $, $\textbf{j}^{\rm T} \cdot \textbf{w} - \frac{1}{2} \textbf{w}^{\rm T} \cdot \textbf{Q} \cdot \textbf{w}$ is very negative, regardless of the value of the other $w_{j \neq i}$. 
    This condition can always be met if we choose a large enough $\Delta^{(i)}$ such that
    \begin{equation}
    \begin{split}
    p(\left\{ C\right\} |w_i = \pm \Delta^{(i)}, H_{\ell^{\rm (inf)}_{\rm max}}) \approx 0. 
    \end{split}
    \end{equation}
\end{enumerate}
This prior corresponds to a square centered at the origin in the complex $\mathcal{P}_{\ell m} $ plane for $(\ell, m) \neq 0$. One may think that a more natural prior would be one that is uniform for, say, $|\mathcal{P}_{\ell m}| \leq \Delta$ with some $\Delta > 0 $, which corresponds to a circle centered at the origin in the complex plane. 
However, if $\Delta$ is large enough, both the square and circle priors will lead to similar parameter estimation results. This is because, in the region between the square and the circle priors, the argument of the exponential in the posterior is very negative, and thus, the contribution to the posterior can be well approximated by zero.
This condition is not contradictory to the weak signal approximation, because it can be met by a smaller $\Delta^{(i)}$, corresponding to a weaker signal if we have more data.

With these properties in place, the marginalized posterior of $w_i$ can be evaluated as 
\begin{widetext}
\begin{align}
p(w_i| \left\{ C\right\}, H_{\ell^{\rm (inf)}_{\rm max}}) & = \int d \tilde{\textbf{w}}^{(i)} p(\textbf{w}| \left\{ C\right\}, H_{\ell^{\rm (inf)}_{\rm max}}) 
\nonumber \\
& = \frac{1}{p(\left\{ C\right\}|H_{\ell^{\rm (inf)}_{\rm max}})} \prod_{j \neq i} \int_{-\Delta^{(i)}}^{\Delta^{(i)}} d w_j  p(\textbf{w}|H) p(\left\{ C\right\} |\textbf{w}, H_{\ell^{\rm (inf)}_{\rm max}}) \nonumber \\
& = \frac{1}{p(\left\{ C\right\}|H_{\ell^{\rm (inf)}_{\rm max}})} \prod_{j \neq i} \int_{-\Delta^{(i)}}^{\Delta^{(i)}} \frac{d w_j}{2 \Delta^{(i)}} p(\left\{ C\right\} |\textbf{w}, H_{\ell^{\rm (inf)}_{\rm max}}) \nonumber \\
& \propto \prod_{j \neq i} \int_{-\Delta^{(i)}}^{\Delta^{(i)}} d w_j \exp \left( j_i w_i - \frac{1}{2} Q_{i i} w_i^2 + {\textbf{n}^{(i)}}{}^{\rm T} \cdot \tilde{\textbf{w}}^{(i)} - \frac{1}{2} {\tilde{\textbf{w}}^{(i)}}{}^{\rm T} \cdot \tilde{\textbf{Q}}^{(i)} \cdot \tilde{\textbf{w}}^{(i)} \right) \nonumber \\
& \approx \prod_{j \neq i} \int_{-\infty}^{+\infty} d w_j \exp \left( j_i w_i - \frac{1}{2} Q_{i i} w_i^2 + {\textbf{n}^{(i)}}{}^{\rm T} \cdot \tilde{\textbf{w}}^{(i)} - \frac{1}{2} {\tilde{\textbf{w}}^{(i)}}{}^{\rm T} \cdot \tilde{\textbf{Q}}^{(i)} \cdot \tilde{\textbf{w}}^{(i)} \right) \nonumber \\
& \propto \exp \left( j_i w_i - \frac{1}{2} Q_{i i} w_i^2 + \frac{1}{2} {\textbf{n}^{(i)}}{}^{\rm T} \cdot \textbf{M}^{(i)} \cdot \textbf{n} \right) \nonumber \\
& = \exp \left( j_i w_i - \frac{1}{2} Q_{i i} w_i^2 + \frac{1}{2} (\textbf{b}^{(i)} - w_i \textbf{a}^{(i)}){}^{\rm T} \cdot \textbf{M}^{(i)} \cdot (\textbf{b}^{(i)} - w_i \textbf{a}^{(i)}) \right) \nonumber \\
& \propto \exp \left[ \left(j_i - {\textbf{a}^{(i)}}{}^{\rm T} \cdot \textbf{M}^{(i)} \cdot \textbf{b}^{(i)}\right) w_i - \frac{1}{2} \left( Q_{i i} - {\textbf{a}^{(i)}}{}^{\rm T} \cdot \textbf{M}^{(i)} \cdot \textbf{a}^{(i)} \right) w_i^2\right], 
\label{eq:main-marginalization-L}
\end{align}
\end{widetext}
where in going from the fourth to the fifth line we have made use of the third property of the uniform prior of $\textbf{w}$, and from the sixth to the seventh line we have used that $\textbf{M}^{(i)}$ is symmetric, so that ${\textbf{b}^{(i)}}{}^{\rm T} \cdot \textbf{M}^{(i)} \cdot \textbf{a}^{(i)} = {\textbf{a}^{(i)}}{}^{\rm T} \cdot \textbf{M}^{(i)} \cdot \textbf{b}^{(i)}$. 
We again remind the reader that the repeated index $i$ does not imply summation.
We see that the marginalized posterior of $w_i$ is a Gaussian function of $w_i$. 
The mean, $\mu_i$, and the standard deviation, $\sigma_{i}$, of $w_i$ can be read from the marginalized posterior of $w_i$ readily, namely
\begin{align}\label{eq:mu_i}
\mu_i & = \frac{j_i - {\textbf{a}^{(i)}}{}^{\rm T} \cdot \textbf{M}^{(i)} \cdot \textbf{b}^{(i)}}{ Q_{i i} - {\textbf{a}^{(i)}}{}^{\rm T} \cdot \textbf{M}^{(i)} \cdot \textbf{a}^{(i)}}, \nonumber \\
\sigma_i & = \left( Q_{i i} - {\textbf{a}^{(i)}}{}^{\rm T} \cdot \textbf{M}^{(i)} \cdot \textbf{a}^{(i)} \right)^{-\frac{1}{2}}.
\end{align}
Note that, throughout the marginalization procedure, we do not ignore the correlations among the components of $w_i$; these correlations are encoded in the off-diagonal elements of $\textbf{Q}$.
If one neglects the correlations among the $w_i$'s, one sets the off-diagonal elements of $\textbf{Q}$ to zero, resulting in a diagonal $\tilde{\textbf{Q}}^{(i)}$ and $\tilde{\textbf{M}}^{(i)}$. 

Equation~\eqref{eq:mu_i} can be more compactly written as 
\begin{equation}\label{eq:mu_i_2}
\begin{split}
\mu_i & = \left[ \textbf{Q}^{-1} \cdot \textbf{j} \right]_i, \\
\sigma_i & = \left[ \textbf{Q}^{-1} \right]_{ii}^{\frac{1}{2}}, \\
\end{split}
\end{equation}
using the inverse formulae of a block matrix. 
To see this, we first consider the case of $i=1$ and write $\textbf{Q}$ as a block matrix and $\textbf{j}$ as 
\begin{equation}
\textbf{Q} = 
\begin{pmatrix}
Q_{11} & {\textbf{a}^{(1)}}^{\rm T}\\
\textbf{a}^{(1)} & \tilde{\textbf{Q}}^{(1)}
\end{pmatrix}, 
~ \textbf{j} = 
\begin{pmatrix}
j_1\\
\textbf{b}^{(1)}
\end{pmatrix}. 
\end{equation}
Then, we compute the inverse of $\textbf{Q}$ using the block-inverse formula. 
For a block matrix $\textbf{P}$ \cite{bernstein2005matrix}, 
\begin{widetext}
\begin{equation}
\begin{split}
\textbf{P}^{-1} & = 
\begin{pmatrix}
\textbf{A} & \textbf{B}\\
\textbf{C} & \textbf{D}
\end{pmatrix}^{-1} = 
\begin{pmatrix}
\left(\mathbf{A}-\mathbf{B D}^{-1} \mathbf{C}\right)^{-1} & -\left(\mathbf{A}-\mathbf{B D}^{-1} \mathbf{C}\right)^{-1} \mathbf{B} \mathbf{D}^{-1} \\
-\mathbf{D}^{-1} \mathbf{C}\left(\mathbf{A}-\mathbf{B D}^{-1} \mathbf{C}\right)^{-1} & \mathbf{D}^{-1}+\mathbf{D}^{-1} \mathbf{C}\left(\mathbf{A}-\mathbf{B D}^{-1} \mathbf{C}\right)^{-1} \mathbf{B} \mathbf{D}^{-1}
\end{pmatrix}. 
\end{split}
\end{equation}
Taking $\textbf{A}=Q_{11}$, $\textbf{B}={\textbf{a}^{(1)}}^{\rm T}$, $\textbf{C}=\textbf{a}^{(1)}$ and $\textbf{D}=\tilde{\textbf{Q}}^{(1)}$, we find 
\begin{equation}
\textbf{Q}^{-1} = 
\begin{pmatrix}
\left(Q_{11}-{\textbf{a}^{(1)}}^{\rm T} \cdot \tilde{\textbf{M}}^{(1)} \cdot \textbf{a}^{(1)}\right)^{-1} & -\left(Q_{11}-{\textbf{a}^{(1)}}^{\rm T} \cdot \tilde{\textbf{M}}^{(1)} \cdot \textbf{a}^{(1)}\right)^{-1} {\textbf{a}^{(1)}}^{\rm T} \tilde{\textbf{M}}^{(1)} \\
-\tilde{\textbf{M}}^{(1)} \cdot \textbf{a}^{(1)} \left(Q_{11}-{\textbf{a}^{(1)}}^{\rm T} \cdot \tilde{\textbf{M}}^{(1)} \cdot \textbf{a}^{(1)}\right)^{-1} & \tilde{\textbf{M}}^{(1)}+ \tilde{\textbf{M}}^{(1)} \cdot \textbf{a}^{(1)} \left(Q_{11}-{\textbf{a}^{(1)}}^{\rm T} \cdot \tilde{\textbf{M}}^{(1)} \cdot \textbf{a}^{(1)}\right)^{-1} {\textbf{a}^{(1)}}^{\rm T} \cdot \tilde{\textbf{M}}^{(1)}. 
\end{pmatrix}
\end{equation}
\end{widetext}
Reading the first row of $\textbf{Q}^{-1} \cdot \textbf{j}$ and the $(1,1)$ element of $\textbf{Q}^{-1}$, we find that 
\begin{equation}
\begin{split}
& \left[ \textbf{Q}^{-1} \cdot \textbf{j} \right]_1 = \frac{j_1 - {\textbf{a}^{(1)}}{}^{\rm T} \cdot \textbf{M}^{(1)} \cdot \textbf{b}^{(1)}}{ Q_{11} - {\textbf{a}^{(1)}}{}^{\rm T} \cdot \textbf{M}^{(1)} \cdot \textbf{a}^{(1)}} = \mu_1, \\
& \left[ \textbf{Q}^{-1} \right]_{11}=\left( Q_{11} - {\textbf{a}^{(1)}}{}^{\rm T} \cdot \textbf{M}^{(1)} \cdot \textbf{a}^{(1)} \right)^{-1}
 = \sigma_1^2. 
\end{split}
\end{equation}
The above arguments can be generalized to other $i \neq 1$. 
A convinient way to generalize the argument is rearranging $\textbf{Q}$ and $\textbf{j}$ into 
\begin{equation}
\begin{split}
\textbf{Q} \rightarrow 
\begin{pmatrix}
Q_{ii} & {\textbf{a}^{(i)}}^{\rm T}\\
\textbf{a}^{(i)} & \tilde{\textbf{Q}}^{(i)}
\end{pmatrix}, 
~ \textbf{j} \rightarrow 
\begin{pmatrix}
j_i\\
\textbf{b}^{(i)}
\end{pmatrix}. 
\end{split}
\end{equation}
Computing $\left[ \textbf{Q}^{-1} \cdot \textbf{j} \right]_1$ and $\textbf{Q}^{-1}$ through the above procedure can then prove the case for $i\neq 1$. 
The marginalization procedure described above, and in particular, Eq.~\eqref{eq:mu_i_2} (or Eq.~\eqref{eq:mu_i}), are some of the key results of this paper. 

We shall conclude this subsection by discussing the relation between our analysis and the Fisher information matrix analysis. 
First, the matrix $\textbf{Q}$ is actually a Fisher information matrix, which can be seen by realizing that
\begin{equation}
Q_{ij} = - \frac{\partial^2}{\partial w_i \partial w_j} \log  p(\textbf{w}| \left\{ C\right\}, H_{\ell^{\rm (inf)}_{\rm max}}). 
\end{equation}
Second, the maximum-likelihood estimation obtained by the Fisher information matrix analysis, which amounts to solving the equations 
\begin{equation}
\frac{\partial}{\partial w_i} \log  p(\textbf{w}| \left\{ C\right\}, H_{\ell^{\rm (inf)}_{\rm max}}) \Bigg|_{\textbf{w} = \textbf{w}_{\rm ML}}= 0 \Rightarrow \textbf{Q} \cdot \textbf{w}_{\rm ML} = \textbf{j}, 
\end{equation}
is actually identitical to the $\mu_i$ given by Eq.~\eqref{eq:mu_i_2}. 
Moreover, following from the usual Fisher information matrix analysis, the measurement uncertainty of $w_i$ is given by the square root of the $(i,i)$ element of the inverse of the Fisher information matrix, 
\begin{equation}
\Delta w_i = \left[ \textbf{Q}^{-1} \right]_{ii}^{\frac{1}{2}}, 
\end{equation}
which is just $\sigma_i$, as given in Eq.~\eqref{eq:mu_i_2}.
In other words, Eq.~\eqref{eq:mu_i_2} recovers exactly the maximum-likelihood estimate and the measurement uncertainty of $\textbf{w}$ obtained using a Fisher information matrix analysis. 
This is reasonable, because $p(\textbf{w}|H_{\ell^{\rm (inf)}_{\rm max}})$ is a constant, which implies that the posterior is proportional to likelihood.
Hence, the maximum-posterior $\textbf{w}$ and maximum-likelihood $\textbf{w}$ are the same, and so is the measurement uncertainty. 
Another consequence of this connection is that $\mu_i$, being the $w_i$ that maximizes the marginalized posterior, is also a component of the maximum-posterior $\textbf{w}$, the latter of which is defined as
\begin{equation}
\textbf{w}_{\rm MP} = \text{arg}\max\limits_{\textbf{w}} p(\textbf{w}|\{ C \}, H_{\ell^{\rm (inf)}_{\rm max}}). 
\end{equation}
Recovering the results obtained through the Fisher information matrix analysis proves the correctness of our marginalization. 

\begin{widetext}

\subsection{Bayes factor}
\label{sec:BF}

Given a large enough $\Delta^{(i)}$, the Bayes factor between the two hypotheses defined in Eq.~\eqref{eq:hypothese} can also be analytically evaluated in a similar manner. 
To calculate the Bayes factor, we need to evaluate $p(\{ C\}|H_{\ell^{\rm (inf)}_{\rm max}})$ and $p(\{ C\}|H_{\rm null})$. 
We first evaluate $p(\{ C\}|H_{\ell^{\rm (inf)}_{\rm max}})$ using Bayes theorem, 
\begin{align} 
& p(\{ C\}|H_{\ell^{\rm (inf)}_{\rm max}}) 
= \int d \textbf{w} \; p(\left\{ C\right\} |\textbf{w}, H_{\ell^{\rm (inf)}_{\rm max}}) \; p(\textbf{w}|H_{\ell^{\rm (inf)}_{\rm max}}) \nonumber \\
& = \bar{\mathcal{N}} \prod_{i=1}^{(\ell_{\max}^{\rm (inf)}+1)^2} \int_{-\Delta^{(i)}}^{\Delta^{(i)}} \frac{d w_i}{2 \Delta^{(i)}} \exp \left( \textbf{j}^{\rm T} \cdot \textbf{w} - \frac{1}{2} \textbf{w}^{\rm T} \cdot \textbf{Q} \cdot \textbf{w} \right) \nonumber \\
& \approx \bar{\mathcal{N}} \prod_{i=1}^{(\ell_{\max}^{\rm (inf)}+1)^2} \int_{-\infty}^{+\infty} \frac{d w_i}{2 \Delta^{(i)}} \exp \left( \textbf{j}^{\rm T} \cdot \textbf{w} - \frac{1}{2} \textbf{w}^{\rm T} \cdot \textbf{Q} \cdot \textbf{w} \right) \nonumber \\
& = \frac{\bar{\mathcal{N}}}{|\textbf{Q}|^{1/2}\Delta(\ell_{\max}^{\rm (inf)})} \left(\frac{\pi}{2} \right)^{\frac{[\ell_{\rm max}^{\rm (inf)}+1]^2}{2}} \exp \left(\frac{1}{2} \textbf{j}^{\rm T} \cdot \textbf{Q}^{-1} \cdot \textbf{j} \right), 
\label{eq:analytic_eval_BF}
\end{align}
\end{widetext}
where from the third to the fourth line we have again made use of the third property of the uniform prior of $\textbf{w}$, $|\textbf{Q}|$ is the determinant of $\textbf{Q}$, and 
\begin{equation}
\Delta(\ell_{\max}^{\rm (inf)}) = \prod_{i = 1}^{[\ell_{\rm max}^{\rm (inf)}+1]^2} \Delta^{(i)}. 
\end{equation}
When the hypothesis is $H_{\rm null}$, the evidence simplifies significantly, as we show below:
\begin{align}
p(\{ C\}|H_{\rm null}) & = \int d \textbf{w} \; p(\left\{ C\right\} |\textbf{w}, H_{\rm null}) \; p(\textbf{w}|H_{\rm null}) \nonumber \\
& = \int d \textbf{w} \; p(\left\{ C\right\} |\textbf{w} = \textbf{0}, H_{\ell^{\rm (inf)}_{\rm max}}) \; p(\textbf{w}|H_{\ell^{\rm (inf)}_{\rm max}}) \nonumber \\
& = p(\left\{ C\right\} |\textbf{w} = \textbf{0}, H_{\ell^{\rm (inf)}_{\rm max}}) \nonumber \\
& = \bar{\mathcal{N}}. 
\end{align}
Thus, the Bayes factor can be analytically evaluated as 
\begin{equation}\label{eq:Bayes_factor}
\begin{split}
& \mathcal{B}(\ell_{\max}^{\rm (inf)}| \{\Delta^{(i)}\}) \\
& = \frac{1}{|\textbf{Q}|^{1/2}\Delta(\ell_{\max}^{\rm (inf)})} \left(\frac{\pi}{2} \right)^{\frac{[\ell_{\rm max}^{\rm (inf)}+1]^2}{2}} \exp \left(\frac{1}{2} \textbf{j}^{\rm T} \cdot \textbf{Q}^{-1} \cdot \textbf{j} \right). 
\end{split}
\end{equation}

At this junction, a word of caution is necessary. 
Equation~\eqref{eq:Bayes_factor} is valid only if a large enough $\Delta^{(i)}$ is chosen, because otherwise one cannot extend the limits of integration in the fifth line of Eq.~\eqref{eq:main-marginalization-L} and in the third equality of Eq.~\eqref{eq:analytic_eval_BF}.  
Apart from this criterion, the width of the prior of $\textbf{w}$ is arbitrary, which means that the Bayes factor is also, in this sense, arbitrary. 
This is because the Bayes factor depends on the prior volume of the parameters that characterize the hypothesis that is being compared. 
Thus, when computing the Bayes factor using Eq.~\eqref{eq:Bayes_factor}, one should also be careful of and report the chosen $\Delta^{(i)}$. 
This is also the reason why the Bayes factor in Eq.~\eqref{eq:Bayes_factor} is written as $\mathcal{B}(\ell_{\max}^{\rm (inf)}|\{ \Delta^{(i)}\})$ to emphasize its dependence on both $\ell_{\rm max}^{\rm (inf)}$ and $\Delta^{(i)}$, both specified according to our hypothesis. 
As we will show in the next section, however, for any reasonably large-enough choice of $\Delta^{(i)}$, the effects of the value of $\Delta^{(i)}$ on the Bayes factor is not significant and will not affect the ranking between the two hypotheses. Therefore, whether we choose $\Delta^{(i)} = 1$ or $\Delta^{(i)} = 10$, both of which are much larger than the astrophysically motivated value of $\mathcal{P}_{\ell m}$, corresponding to $|\mathcal{P}_{\ell m}| \sim \mathcal{O}(10^{-48})$, our conclusions will be unaffected. 

Let us conclude this section by pointing out that the above calculations can be easily extended to a detector network that contains more detectors. 
To apply the method to a detector network, one just sums over the detector pairs when calculating the following quantities \cite{Mitra:2007mc}
\begin{equation}
\begin{split}
\textbf{j} & = \sum_{I} \sum_{J>I} \textbf{j}^{(IJ)}, \\
\textbf{Q} & = \sum_{I} \sum_{J>I} \textbf{Q}^{(IJ)}, 
\end{split}
\end{equation}
where $\textbf{j}^{(IJ)}$ and $\textbf{Q}^{(IJ)}$ are respectively the $j$ vector and $\textbf{Q}$ of the detectors $I$ and $J$ (c.f. Eq.~\eqref{eq:j_and_Q}). 

\section{Mock data analysis}
\label{sec:Mock_data}

In this section, we illustrate the accuracy of our analysis in extracting the angular structures of a GWB by applying it to mock data. 
We will first explain the general setup of different mock data analyses. 
Then, we will apply our analysis to different sets of mock data, each corresponding to a different level of anisotropy. 
We will show that our analysis can extract the angular structure of different types of anisotropic sources with excellent accuracy. 

\subsection{General setup}

As the likelihood (Eq.~\eqref{eq:likelihood}) does not explicitly depend on the strain data measured by individual detectors but on their correlation, we follow \cite{Tsukada:2022nsu} and directly simulate the cross-spectral density of data segments in the frequency domain, 
\begin{equation}
\label{eq:cross-spectral-inj}
C_{\rm inj}(f, t) = C_{n}(f, t) + H_{\alpha} (f) \sum_{\ell=0}^{\ell^{\rm (inj)}_{\max}} \sum_{m=-\ell}^{\ell} \gamma_{\ell m}(f, t) \mathcal{P}^{\rm (inj)}_{\ell m}.
\end{equation}
In this expression, $\ell^{\rm (inj)}_{\max}$ and $\mathcal{P}^{\rm (inj)}_{\ell m}$ are the maximum $\ell$ and the spherical-harmonic components of the simulated GWB contained in the mock data respectively. 
Note that $\ell^{\rm (inj)}_{\max}$ is in general different from $\ell_{\rm max}^{\rm (inf)}$ because the maximum $\ell$ that a GWB corresponds to can, in general, be different from the maximum $\ell$ that we choose to infer. 
Through the mock data analyses, we choose $\ell^{\rm (inf)}_{\max} = 1, 2, ..., 10$, but in general $\ell^{\rm (inf)}_{\rm max}$ can be freely adjusted for analyzing actual data. 
Note also that by directly simulating the cross-spectral density in the frequency domain, we are assuming A.4 and ignoring the cross- and auto-correlations present in the time-domain data. 
In practice, the analysis should thus be applied to individual (windowed) segments and then optimally combine the result from individual segments to address the cross- and auto-correlations. 
Nonetheless, as pointed out when assuming that condition A.4 holds; if the windowing and optimal combinations are properly executed, the final results should agree well with calculations that use the likelihood and ignores these correlations, as shown in \cite{Dipole_01}. 

We study the effects of noise fluctuations by including $C_{n}(f, t)$ in the injected cross-spectral density of data segments in Eq.~\eqref{eq:cross-spectral-inj}. In particular, $C_{n}(f, t)$  represents the cross-spectral density of the stationary Gaussian noise contained in the data. We simulate $C_n(f, t)$ by generating a random complex frequency sequence of zero mean and variance that satisfies \cite{Tsukada:2022nsu}
\begin{equation}
\left\langle|C_n (f, t)|^2\right\rangle-|\langle C_n (f, t)\rangle|^2 \approx \frac{N^{\rm (n)}_I(f,t) N^{\rm (n)}_J(f,t)}{\tau \Delta f}, 
\end{equation}
where recall that $\tau$ is the length of the data segments, $\Delta f$ is the frequency resolution and $N^{\rm (n)}_{I, J}(f,t)$ are the noise PSDs of the detectors $I$ and $J$, respectively. 
Since the measured strain data contain both the instrumental noise and the signal when a GWB presents, the PSD of the strain data measured by individual detectors will contain both the instrumental-noise PSD, $N^{\rm (n)}_{I, J}(f, t)$, and the autocorrelated power of the responses due to a GWB, 
\begin{equation}
N_{I, J}(f, t) = N^{\rm (n)}_{I, J}(f, t) + S_{h} (f, t). 
\end{equation}
Hence, in practice, this $N^{(n)}_{I, J}$ is not the same as the PSD in Eq.~\eqref{eq:likelihood}. 
These two PSDs are extremely difficult to separate in an actual detection. 
Since we expect the signal to be weak, we just regard the measured strain PSD as the noise PSDs for the evaluation of the likelihood, at the cost of slightly reducing our search sensitivity \cite{Biscoveanu:2020gds}. 
To account for this effect, in our mock-data analyses we include both PSDs in our search when evaluating the likelihood and marginalized posteriors, but we only include $N_{I, J}^{(n)}(f, t)$ when simulating $C_n (f, t)$. 

\begin{figure*}[tp!]
\centering  
\subfloat{\includegraphics[width=0.47\linewidth]{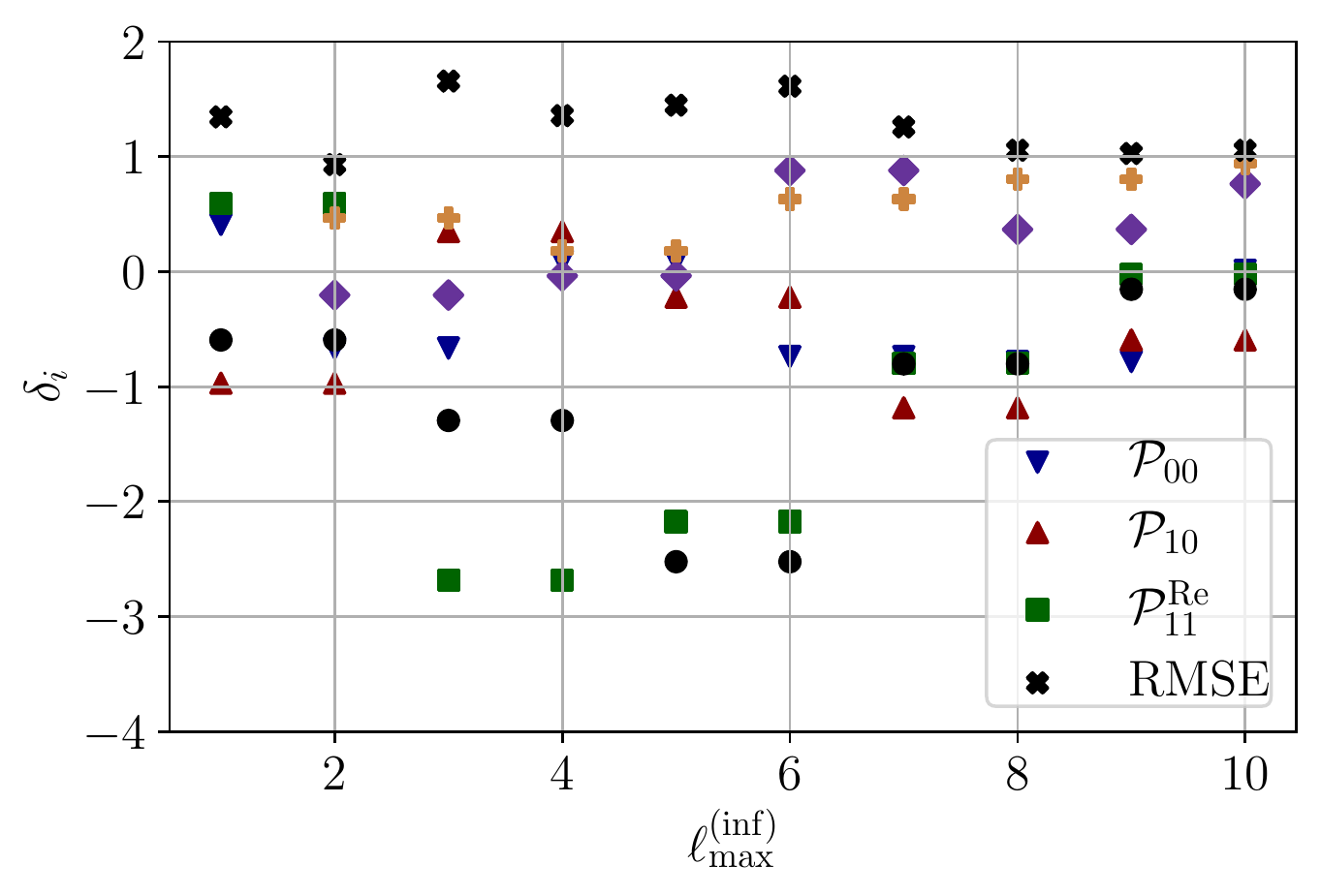}}
\subfloat{\includegraphics[width=0.47\linewidth]{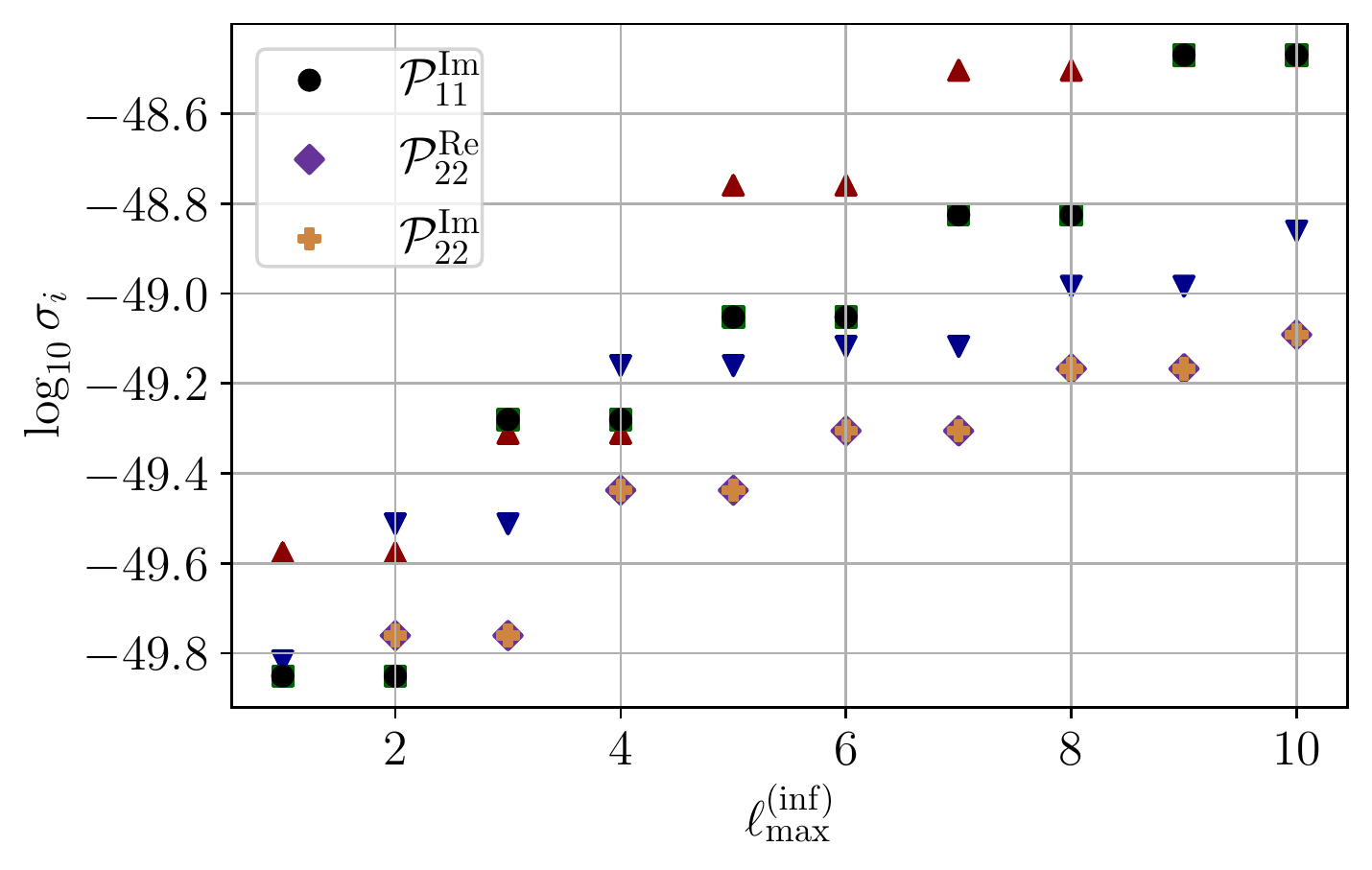}}
\caption{The measurement bias $\delta_i$ (left, see Eq.~\eqref{eq:rho} in the main text) and measurement uncertainty $\sigma_i$ (right) of some $\mathcal{P}_{\ell m}$, obtained by applying our analysis to one year of mock data, which contain solely stationary Gaussian noise, as a function of $\ell^{\rm (inf)}_{\rm max}$, assuming $\alpha=2/3$. 
Note that $\delta_i$ has been scaled by $\sigma_i$ in its definition. 
For the purpose of illustration, we only show $\delta_i$ and $\sigma_i$ for $\mathcal{P}_{00}, \mathcal{P}_{10}, \mathcal{P}_{11}$ and $\mathcal{P}_{22}$. 
Observe that $|\delta_i| < 3$ for all $\mathcal{P}_{\ell m}$, indicating that the results are consistent with the fact that the mock data contain no signal to $3\sigma_i$ confidence. 
Observe also that, as $\ell^{\rm (inf)}_{\rm max}$ approaches 10, $\delta_{\rm RMSE}$ (see Eq.~\eqref{eq:RMSE} for definition) approaches to $\sim 1$ steadily, indicating that overall, the recovered $w_i$ is consistent with zero within $\sim 1\sigma$. }
 \label{fig:Noise_results}
\end{figure*}

Other properties of the injection are chosen to remain in line with current GWB searches with advanced LIGO and Virgo detectors \cite{LIGOScientific:2019gaw, KAGRA:2021mth}. 
More specifically, for each mock analysis, we simulate data that consist of segments of equal time length $\tau = 192 \rm s$. 
Since these mock data analyses are meant to represent proof-of-principle demonstrations, we only simulate data measured by the advanced LIGO Hanford and Livingston detectors at their design sensitivity. 
The PSD of the detectors is estimated with the exact frequency resolution of the cross-spectral density segments to avoid the need for coarse-graining data \cite{LIGO_SGWB_01, LIGO_SGWB_02, LIGO_SGWB_03, LIGO_SGWB_04}. 
As the mock data contain only stationary Gaussian noise and the responses induced by the simulated stationary GWB, we drop the time dependence of the PSDs, so that $N_{I, J}(f, t) = N_{I, J}(f)$ and we do not notch the data at particular frequency bins. 
We assume the data start at the starting time of the third observing run of the advanced LIGO and Virgo detectors.  
We also focus on simulating and searching for GWB with $\alpha = 0$, $2/3$, and $3$ because GWBs characterized by these $\alpha$ are under extensive search and correspond to astrophysical interesting sources. 
More explicitly, $\alpha = 0$ describes the GWB produced by cosmic strings formed during the end of cosmological inflation \cite{Caprini:2018mtu, Guzzetti:2016mkm, Contaldi:2018akz, Easther:2006vd, 1979JETPL..30..682S, Cook:2011hg, Turner:1996ck, Easther:2006gt, Romero:2021kby, 10.21468/SciPostPhysLectNotes.24, Weir:2017wfa, Caprini:2015zlo, Kahniashvili:2008pe, Kahniashvili:2009mf, Caprini:2009yp, Kisslinger:2015hua, RoperPol:2019wvy, Mairi_CS_SGWB_01, Mairi_CS_SGWB_02, Mairi_CS_SGWB_03, SGWB_CS_01, SGWB_CS_02, LIGOScientific:2017ikf, Jenkins:2018nty, LIGOScientific:2021nrg, SGWB_CS_03}. The spectral tilt $\alpha = 2/3$ characterizes the GWB produced by CBCs \cite{Zhu:2012xw, Zhao:2020iew, 10.1111/j.1365-2966.2003.07176.x, Capurri:2021zli}, and $\alpha = 3$ approximately describe the GWB produced by supernova \cite{SGWB_CCSN_01, SGWB_CCSN_02, SGWB_CCSN_03, SGWB_CCSN_04}. 
The explicit value of the spherical-harmonic components of the simulated GWBs will be given individually in the corresponding sections below.

To gauge the accuracy of measuring $w_i$ from the simulated data, for different $i$, we define two measures. 
The first measure is the error of a specific $w_i$ relative to $\sigma_i$,
\begin{equation}\label{eq:rho}
\delta_i = \frac{\mu_i - w_i^{\rm (inj)}}{\sigma_i}\,,
\end{equation}
where recall that $\mu_i$ is the mean of the marginalized posterior of $w_i$, while $\sigma_i$ is its variance.
By examining $\delta_i$, we can study the effects of $\ell^{\rm (inf)}_{\rm max}$ on measurement accuracy of $w_i$. 
If $\delta_i = N$, then the best-fit $w_i$ is $N\sigma$ away from the injected value. Therefore, when $\delta_i$ is close to zero, then the recovered $w_i$ is perfectly consistent with the injected $w_i^{\rm (inj)}$. 
However, due to the presence of noise fluctuations, we expect that $|\delta_i|$ can occasionally be as large as $\sim 3$ (see e.g.~\cite{KAGRA:2021mth}, where the SNR of a GWB is 3.6, but one still cannot claim a detection). 
In what follows, we calculate the marginalized posterior of many parameters, but we will only show results (e.g.~$\delta_i$ and $\sigma_i$) for a subset of them. 
In the Supplementary Material, we present all results obtained by our mock data analysis.

The second measure is the root mean square error (RMSE), 
\begin{equation}\label{eq:RMSE}
\delta_{\rm RMSE} = \sqrt{\frac{1}{(\ell^{\rm (inf)}_{\rm}+1)^2}\sum_{i} \sigma_i^2}, 
\end{equation}
where $\sum_i$ stands for summation over the index $i$ that labels the vector $w_i$, corresponding to $\mathcal{P}_{\ell m}$ for $0 \leq \ell \leq \ell^{\rm (inf)}_{\rm max}$. 
Heuristically, $\delta_{\rm RMSE}$ gives the averaged deviation of the measured $w_i$ from the simulated $w_i$ relative to $\sigma_i$. 
Therefore, unlike $\delta_i$, $\delta_{\rm RMSE}$ measures the overall accuracy of all $w_i$. 

\subsection{Pure-noise injection}

\begin{figure}[tp!]
\includegraphics[width=\columnwidth]{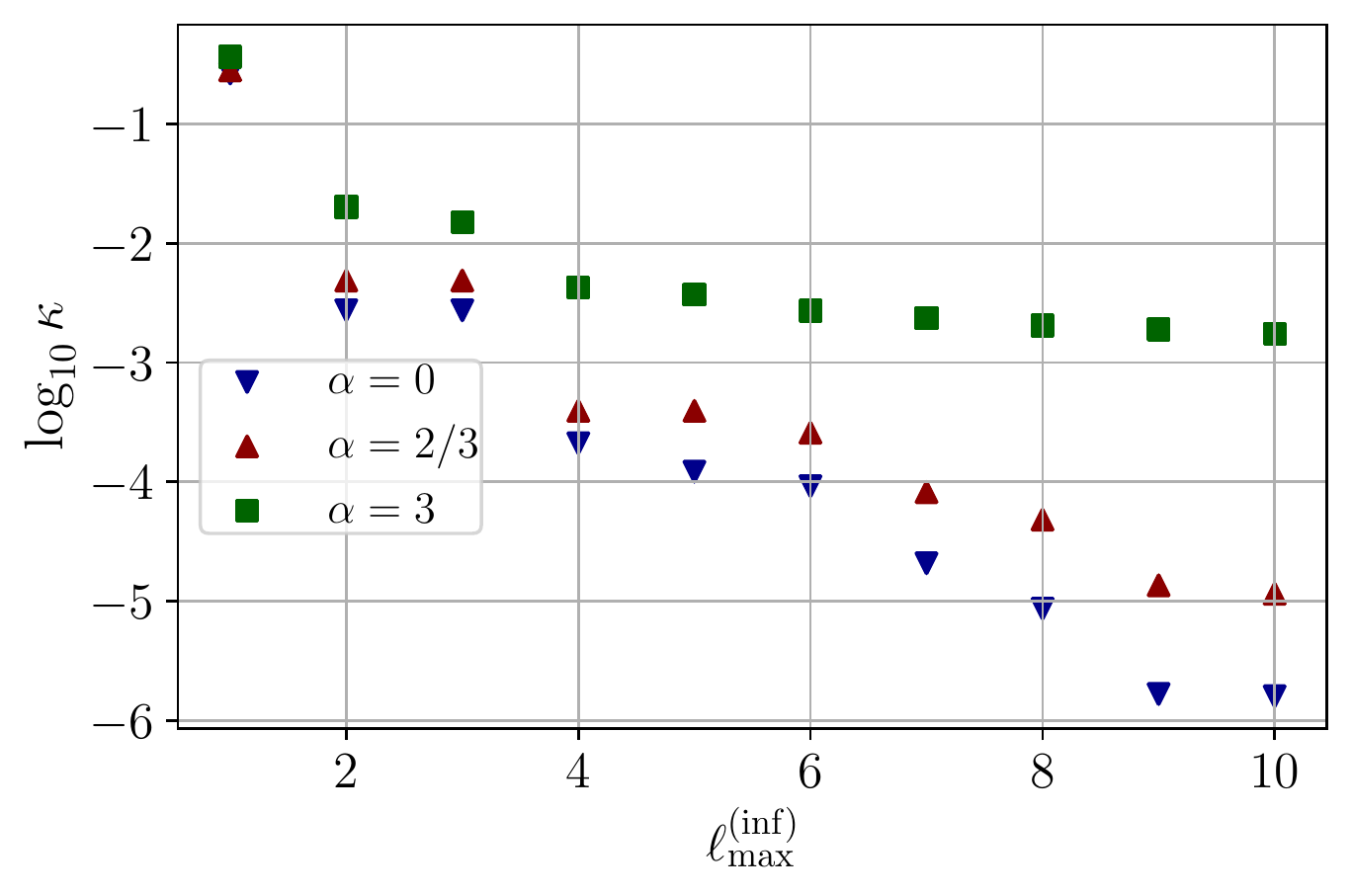}
\caption{The overall condition number ($\kappa$, defined by Eq.~\ref{eq:overall_condition_nr}) of $\tilde{\textbf{Q}}^{(i)}$ (defined below Eq.~\ref{eq:exponent}), whose inverse is required for the evaluation of the marginalized posterior, as a function of $\ell_{\rm max}^{\rm (inf)}$ for $\alpha = 0, 2/3$, and 3. 
A larger $\kappa$ implies that $\tilde{\textbf{Q}}^{(i)}$ is more numerically invertible for all $i$. 
Observe that $\kappa > 10^{-6} $ for $\ell_{\rm max}^{\rm (inf)} = 10$ and all $\alpha$, implying that $\tilde{\textbf{Q}}^{(i)}$ for all $i$ and $\alpha$ can be inverted within double precision. 
}
\label{fig:ConditionNr}
\end{figure}

\begin{figure*}[htp!]
\centering  
\subfloat{\includegraphics[width=0.47\linewidth]{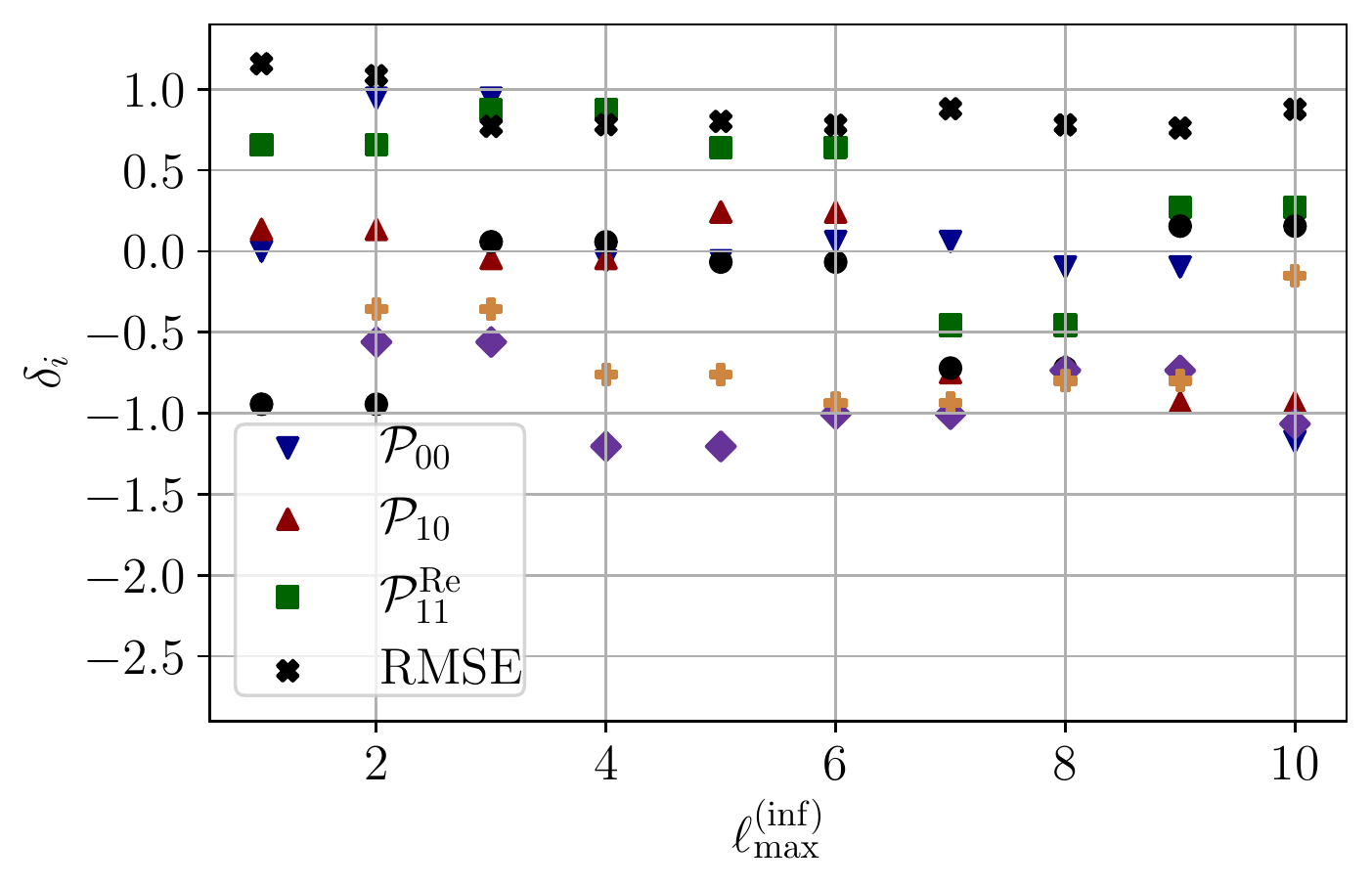}}
\subfloat{\includegraphics[width=0.47\linewidth]{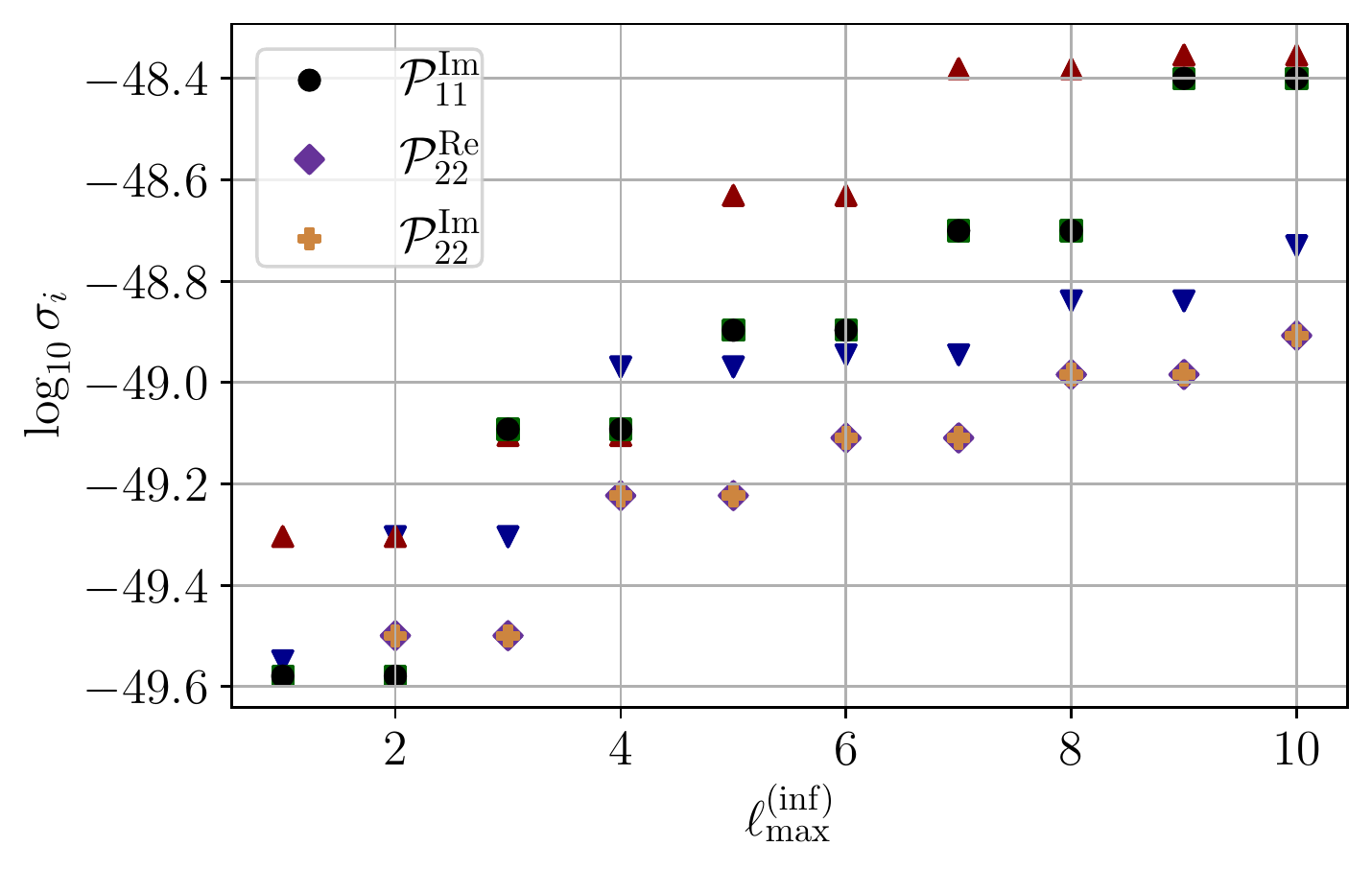}}
\caption{Same as Fig.~\ref{fig:Noise_results} but with mock data that contains a simulated time-independent dipole (i.e. $\mathcal{P}_{\ell \geq 2 m} \equiv 0 $). 
Observe that $|\delta_i| < 3$ for different $\mathcal{P}_{\ell m}$, indicating that our analyses can accurately measure different $\mathcal{P}_{\ell m}$ to $3\sigma$ confidence. 
Moreover, comparing the $\delta_i$ and $\sigma_i$ of $\mathcal{P}_{10}$ and $\mathcal{P}_{11}$ with those of $\mathcal{P}_{22}$ suggests that our analysis is unlikely to mistake a spherical-harmonic component of $\ell \leq \ell_{\rm max}^{\rm (inj)}$ with another spherical-harmonic component of of $\ell > \ell_{\rm max}^{\rm (inj)}$.  
Observe also that, as $\ell^{\rm (inf)}_{\rm max}$ approaches 10, $\delta_{\rm RMSE}$ approaches $1$ steadily, indicating that overall, the recovered $w_i$ is consistent with $w^{\rm (inj)}_i$ within $\sim 1\sigma$. 
}
 \label{fig:TIDipole_results}
\end{figure*}

We first apply our formalism to 365 days of mock data that contain only pure noise. 
The left panel of Fig.~\ref{fig:Noise_results} shows $\delta_i$ and the right panel the base-10 logarithm of $\sigma_i$, both as a function of $\ell^{\rm (inf)}_{\rm max}$. 
To illustrate, we show only the marginalized posterior of $\mathcal{P}_{00}, \mathcal{P}_{10}, \mathcal{P}_{11}^{\rm Re}, \mathcal{P}_{11}^{\rm Im}, \mathcal{P}_{22}^{\rm Re}$ and $\mathcal{P}_{22}^{\rm Im}$. 
Since the results of different $\alpha$ are quantitatively the same, for illustration, we only show $\alpha = 2/3$, corresponding to the GWB formed by CBCs. 
First, we observe that for all $\ell^{\rm (inf)}_{\rm max}$, $|\delta_i|<3$; this means that $\mu_i$ is consistent with $w_i^{\rm (inj)} = 0$ to $3\sigma_i$, indicating that we can accurately pinpoint the fact that the mock data contain no GWB. 
Second, we observe that $\sigma_i$ increases with $\ell^{\rm (inj)}_{\rm max}$ and it is expected. 
Increasing $\ell^{\rm (inj)}_{\rm max}$ introduces more (unnecessary) free parameters whose measure uncertainty correlates with those associated with the spherical-harmonic components of smaller $\ell$, deteriorating the overall measurement accuracy. 

We also check that $\tilde{\textbf{Q}}^{(i)}$ is numerically well conditioned because the evaluation of $\mu_i$ and $\sigma_i$ involves the inverse of $\tilde{\textbf{Q}}^{(i)}$. 
To this end, we compute the individual condition number $\kappa_i$ of the matrix $\tilde{\textbf{Q}}^{(i)}$, which is defined by \footnote{This is not the usual definition of the condition number of a matrix, which is defined as the ratio between the eigenvalue of the largest modulus and that of the least modulus. 
The definition in this paper follows the convention in the literature of the search of anisotropic GWB, e.g., \cite{Thrane:2009fp} and \cite{SHD_01}. } 
\begin{equation}
\kappa_i = \frac{\lambda^{\min}_i}{\lambda^{\max}_i}, 
\end{equation}
where $\lambda^{\min}_i$ and $\lambda^{\max}_i$ are the eigenvalues of $\tilde{\textbf{Q}}^{(i)}$ that have the smallest and largest modulus, respectively.
A larger $\kappa_i$ implies that $\tilde{\textbf{Q}}^{(i)}$ is easier to invert numerically and $\kappa_i = 0 $ means that $\tilde{\textbf{Q}}^{(i)}$ is singular. 
Then, we define the overall condition number $\kappa$ as 
\begin{equation}\label{eq:overall_condition_nr}
\kappa = \min\limits_i \kappa_i. 
\end{equation}
Since $\kappa$ is essentially the lower bound of $\kappa_i$, a larger $\kappa$ implies that $\tilde{\textbf{Q}}^{(i)}$ is easier to numerically invert for all $i$. 
Figure~\ref{fig:ConditionNr} shows $\kappa$ as a function of $\ell_{\rm max}^{\rm (inf)}$ for $\alpha = 0, 2/3$ and $3$. 
Observe that, for $\alpha = 0, 2/3$ and $3$, $\kappa > 10^{-6} $ for $\ell_{\rm max}^{\rm (inf)} = 10$, the upper limit of $\ell_{\rm max}$ considered throughout the paper. This means that $\tilde{\textbf{Q}}^{(i)}$ for all $i$ and $\alpha$ can be inverted within double precision without numerical issues\footnote{In principle, a regularization scheme, such as that presented in \cite{Thrane:2009fp, Romano:2016dpx, Panda:2019hyg}, can also be applied when inverting $\tilde{\textbf{Q}}^{(i)}$, but such regularization may bias results \cite{Ballmer:2005uw, Thrane:2009fp, Suresh:2020khz, Agarwal:2021gvz, Floden:2022scq, Agarwal:2023lzz}.}.  
To further check that $\tilde{\textbf{Q}}^{(i)}$ is properly inverted, we compute the max norm, the maximum of the modulus of the elements of a matrix, of the following error matrix, 
\begin{equation}
E^{(i)} = I - \tilde{\textbf{Q}}^{(i)} \textbf{M}^{(i)}, 
\end{equation}
which should be a zero matrix if $\textbf{M}^{(i)}$ is exactly equal to the inverse of $\tilde{\textbf{Q}}^{(i)}$. 
We find that the max norm of $E_i$ is at most $10^{-10}$ for different $i$ and $\alpha$, confirming that $\tilde{\textbf{Q}}^{(i)}$ can be inverted within double precision without numerical issues. 

\begin{figure*}[tp!]
\centering  
\subfloat{\includegraphics[width=0.47\linewidth]{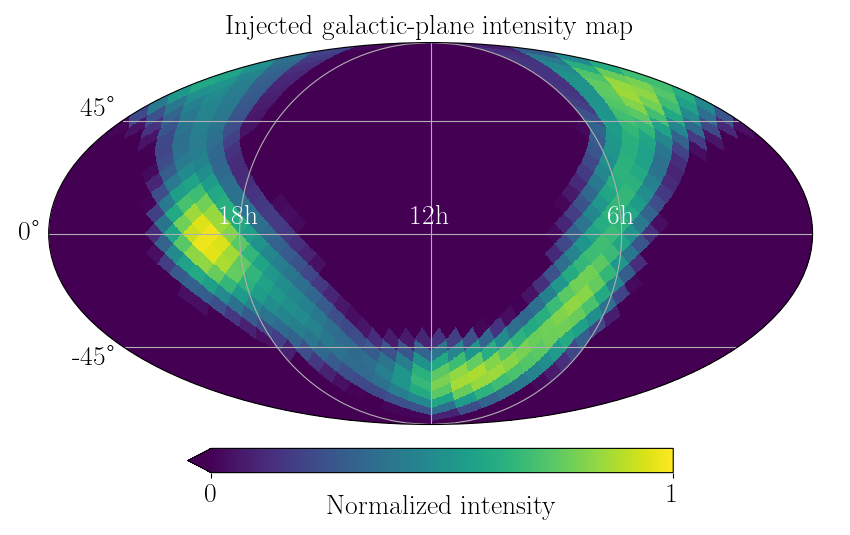}}
\subfloat{\includegraphics[width=0.47\linewidth]{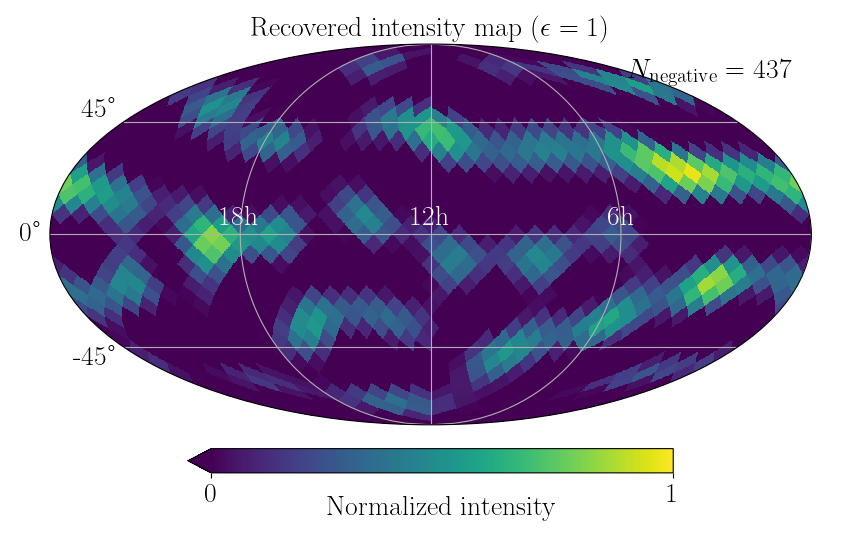}}
\qquad
\subfloat{\includegraphics[width=0.47\linewidth]{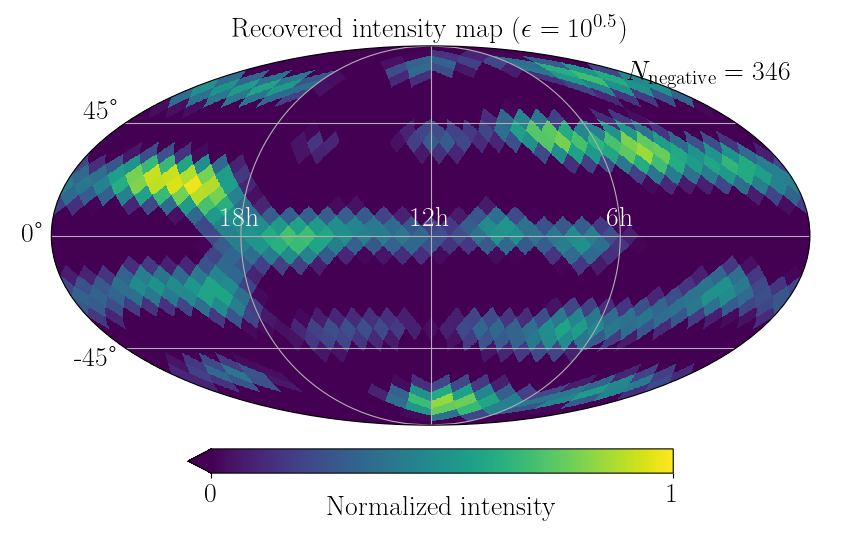}}
\subfloat{\includegraphics[width=0.47\linewidth]{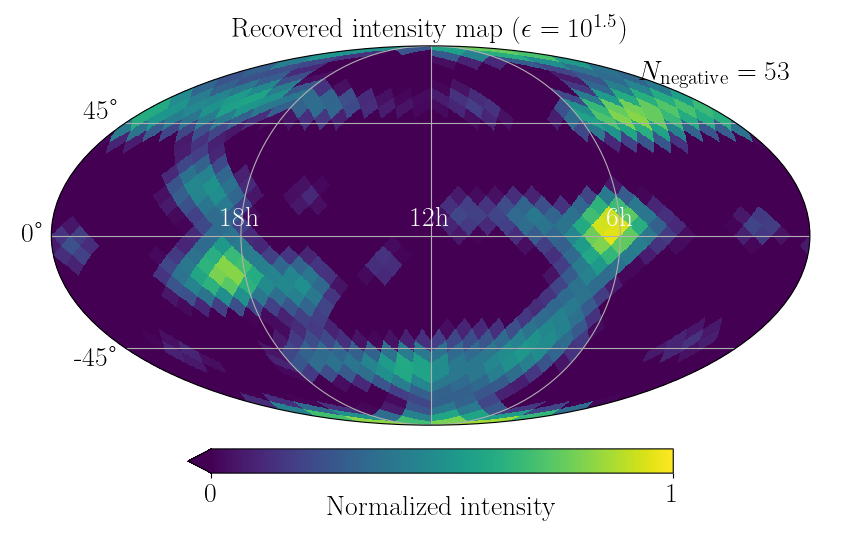}}
\qquad
\subfloat{\includegraphics[width=0.47\linewidth]{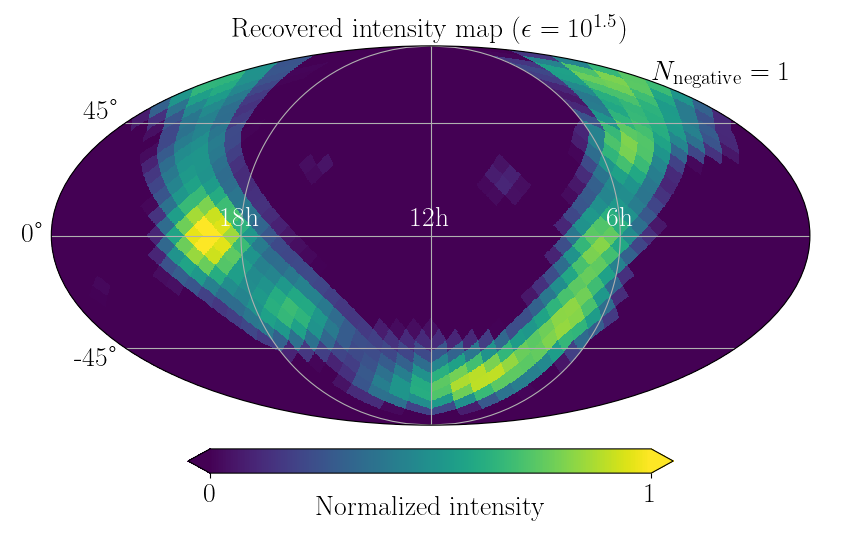}}
\subfloat{\includegraphics[width=0.47\linewidth]{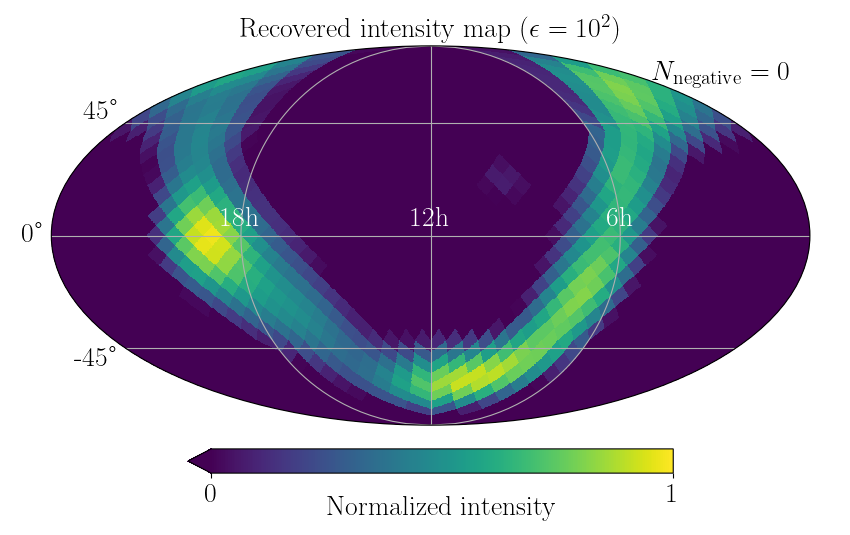}}
\caption{The top, left panel shows the angular distribution of gravitational-wave backgrounds produced by sources populating the galactic plane, which we simulate assuming $\alpha = 2/3$ and that the signal lasts for 30 sidereal days.
The rest of the figures are the recovered intensity maps from mock data containing signals of different strengths, as characterized by $\epsilon$ (see Eq.~(\ref{eq:Plm_galactic_plane}) for definition). 
The intensity maps are presented in the equatorial coordinate system. 
The brightest spot (on the left) is the galactic center. 
All figures are visualized by Mollweide projections and contain 1200 pixels ($N_{\rm side} = 10$).
The number $N_{\rm negative}$ at the top right corner is the number of pixel of the recovered map which has negative intensity. 
The signal-to-noise ratio of the monopole of the background when $\epsilon = 1$ is 15.9 and that when $\epsilon = 10^2 $ is $1050$.  
To show the intensity contrast across different sky directions, the brightness of the color in all panels represents the intensity, and the intensity of all maps is scaled by a number such that the maximum intensity of the simulated map is normalized to one. 
Observe that, as the signal-to-noise ratio of the gravitational-wave background increases, our analysis can recover an intensity map that is increasingly accurate and consistent with the simulated angular distribution.  
Moreover, when the signal-to-noise ratio of the monopole part of the background has reached $\sim 10^3$, the reconstructed intensity maps show almost no visual difference relative to the simulated map. 
This close consistency shows that our formalism is capable of resolving detailed and sophisticated angular structures of a gravitational-wave background. 
}
\label{fig:galactic_plane}
\end{figure*}

\begin{figure}[tp!]
\includegraphics[width=\columnwidth]{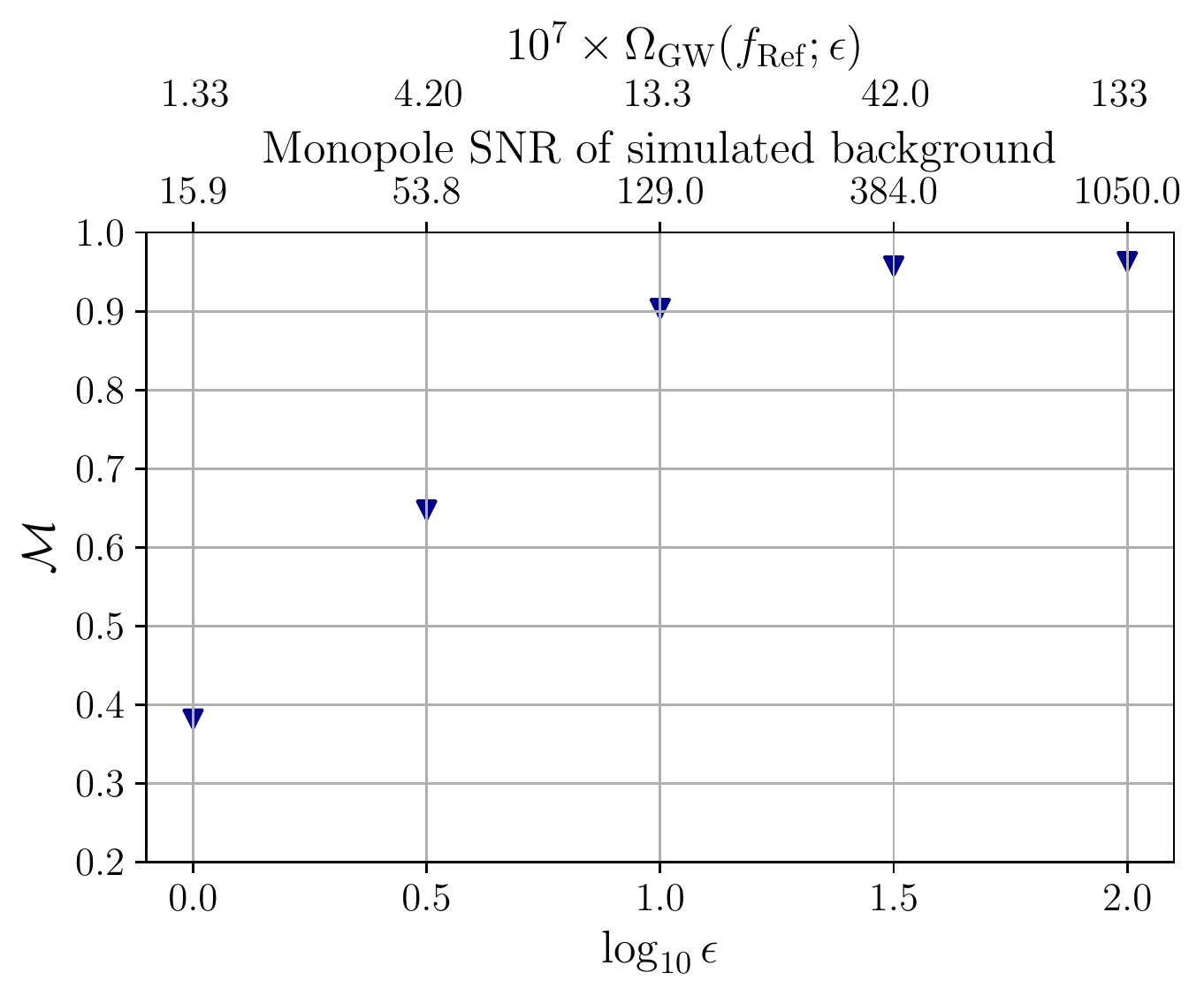}
\caption{
The match between the simulated intensity map of the gravitational-wave background and that recovered using the untargeted Bayesian search (see Eq.~(\ref{eq:match}) for the definition). 
The lower horizontal axis represents the base-10 logarithm of $\epsilon$, a proportionality constant that regulates the amplitude of the simulated gravitational-wave background. 
The upper horizontal axis represents the signal-to-noise ratio of the monopole of the simulated background of the corresponding $\epsilon$. 
On top of the signal-to-noise ratio, we also include the value of $\Omega_{\rm GW}(f; \epsilon)$ at the corresponding $\epsilon$ for reference. 
A match closer to one indicates a more faithful recovery of the intensity map of the gravitational-wave background using our analysis. 
Observe that as $\epsilon$ increases (or equivalently, as the signal-to-noise ratio increases), $\mathcal{M}$ also increases, showing that the recovered intensity map is increasingly accurate for louder signals. 
Moreover, the match is close to one when the monopole signal-to-noise ratio is about 400, which can be achieved within approximately a year using next-generation detectors if a gravitational-wave background of $\Omega_{\rm GW}(f_{\rm Ref}) \sim 10^{-10}$ is present. 
This suggests that our untargeted Bayesian search can indeed be applied to actual gravitational-wave detection in the future. 
}
\label{fig:Match}
\end{figure}

\subsection{Time-independent dipole}
\label{sec:injection_TID}

We now validate our method by recovering a simulated time-independent dipole with $\alpha = 0, 2/3$ and $3$ from 365 days of mock data. 
The simulated dipole signals are motivated by the dipole produced by the peculiar motion of the Solar System barycenter relative to the cosmic rest frame  \footnote{
The orbit of the Earth around the Solar System barycenter induces a smaller time-dependent kinematic dipole signal, even though it is routine to analyze the stochastic GWB from the perspective of the solar system barycenter. This requires special approaches to extract \cite{Dipole_01, Dipole_02, Dipole_03}. 
}. 
For all $\alpha$, the nonzero spherical-harmonic components from the mock data injections are
\begin{equation}
\begin{split}
& \mathcal{P}^{\rm (inj)}_{00} = 4.69 \times 10^{-46}, \\
& \mathcal{P}^{\rm (inj)}_{10} = -1.16 \times 10^{-47}, \\
& \mathcal{P}^{\rm (inj)}_{11} = (6.60+1.41i) \times 10^{-47}, 
\end{split}
\end{equation}
and $\ell^{\rm (inj)}_{\rm max} = 1$. 
These spherical-harmonic components are chosen so that their value is significantly larger than the corresponding measurement uncertainty, facilitating the validation of our analysis. 
The monopole signal is included so that the intensity map is positive in all sky directions. 

Figure~\ref{fig:TIDipole_results} shows $\delta_i$ and $\sigma_i$ for $\mathcal{P}_{00}, \mathcal{P}_{10}, \mathcal{P}_{11}^{\rm Re}, \mathcal{P}_{11}^{\rm Im}, \mathcal{P}_{22}^{\rm Re}$ and $\mathcal{P}_{22}^{\rm Im}$ with $\alpha = 2/3$, obtained by analyzing the mock data with the simulated dipole signal. 
Observe that $|\delta_i| < 3 $ for different $\ell^{\rm (inf)}_{\rm max}$, which shows the robustness of our analysis in two ways. 
First, our analysis can correctly infer different $\mathcal{P}_{\ell m}$ to $3\sigma_i$ confidence. 
In other words, our analysis does not mistake the angular structure of $ \ell \leq \ell^{\rm (inj)}_{\rm max}$ with the angular structure of $\ell^{\rm (inj)}_{\rm max} < \ell \leq \ell^{\rm (inf)}_{\rm max} $. 
Second, choosing different $\ell^{\rm (inf)}_{\rm max}$ does not significantly affect our measurement of $\mathcal{P}_{\ell m}^{\rm (inj)}$. 
Thus, one can adjust $\ell^{\rm (inj)}_{\rm max}$ for the search of different GWB without having to worry that the results will be significantly affected by this choice.
Note that the measurement uncertainty for different $i$ is slightly larger than those shown in Fig.~\ref{fig:Noise_results} due to the contribution of the detectors' PSD from the monopole of the simulated GWB. 

\subsection{Galactic plane distribution}
\label{sec:injection_galactic}

\begin{figure*}[tp!]
\centering  
\subfloat{\includegraphics[width=0.47\linewidth]{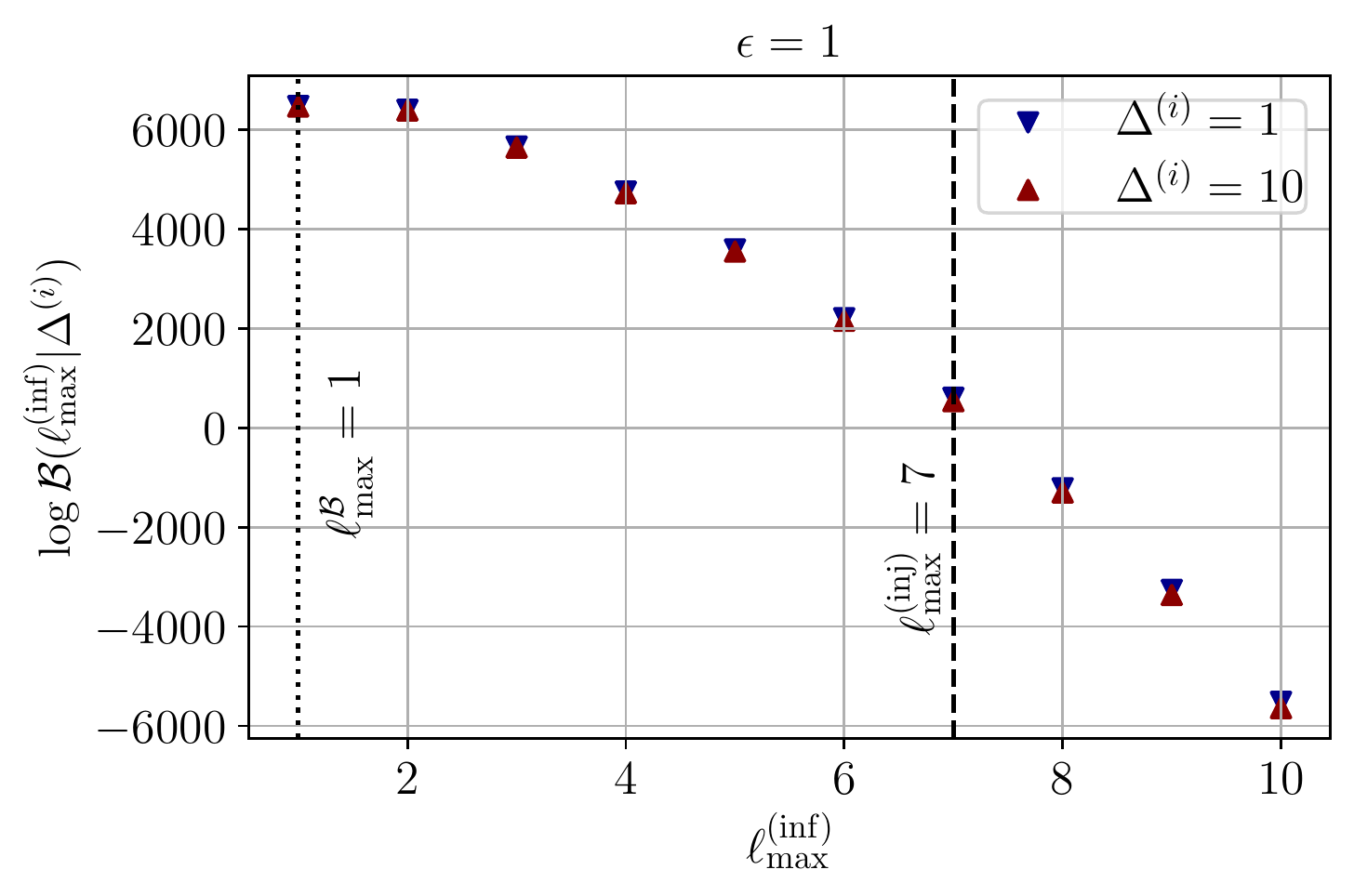}}
\subfloat{\includegraphics[width=0.47\linewidth]{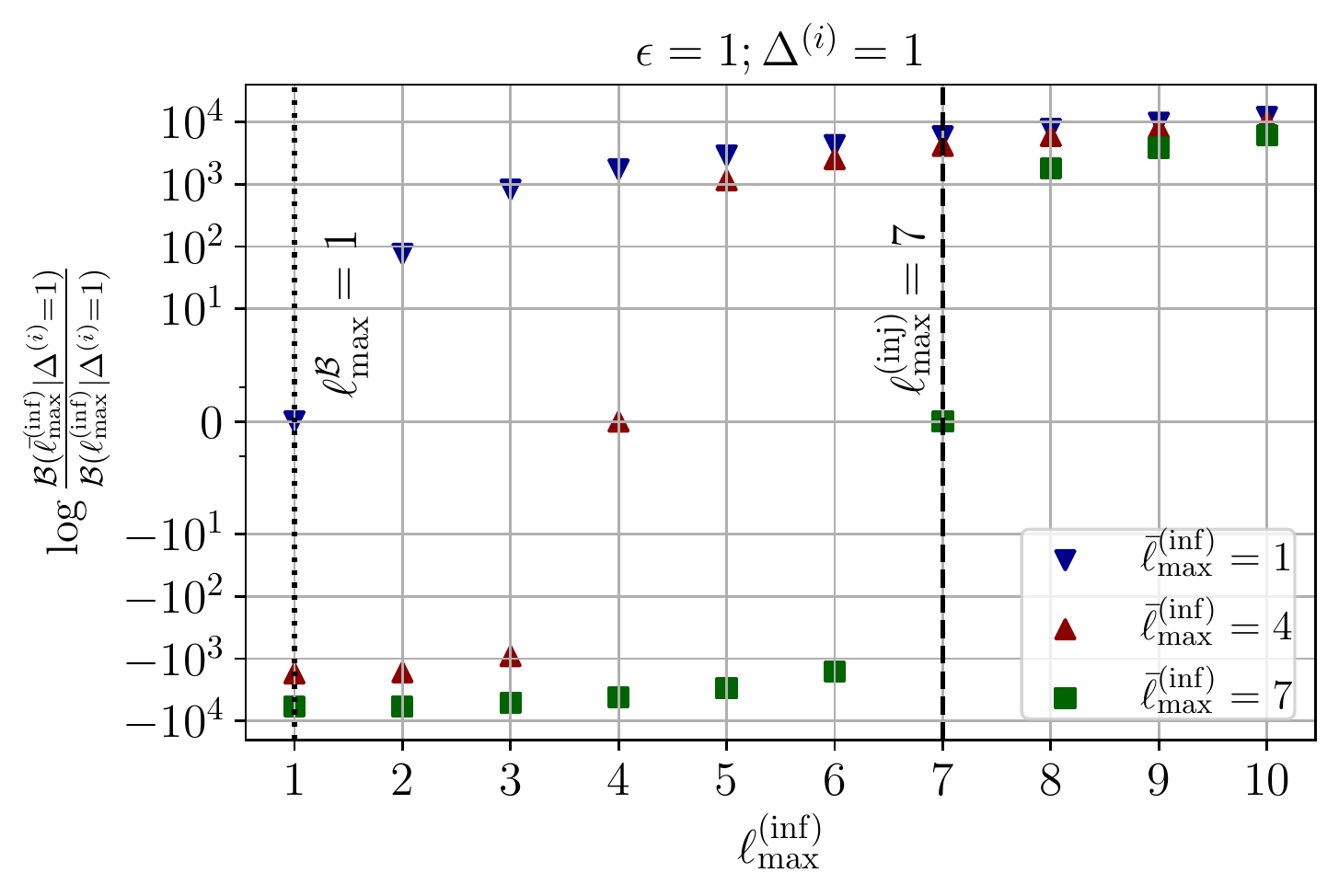}}
\qquad
\subfloat{\includegraphics[width=0.47\linewidth]{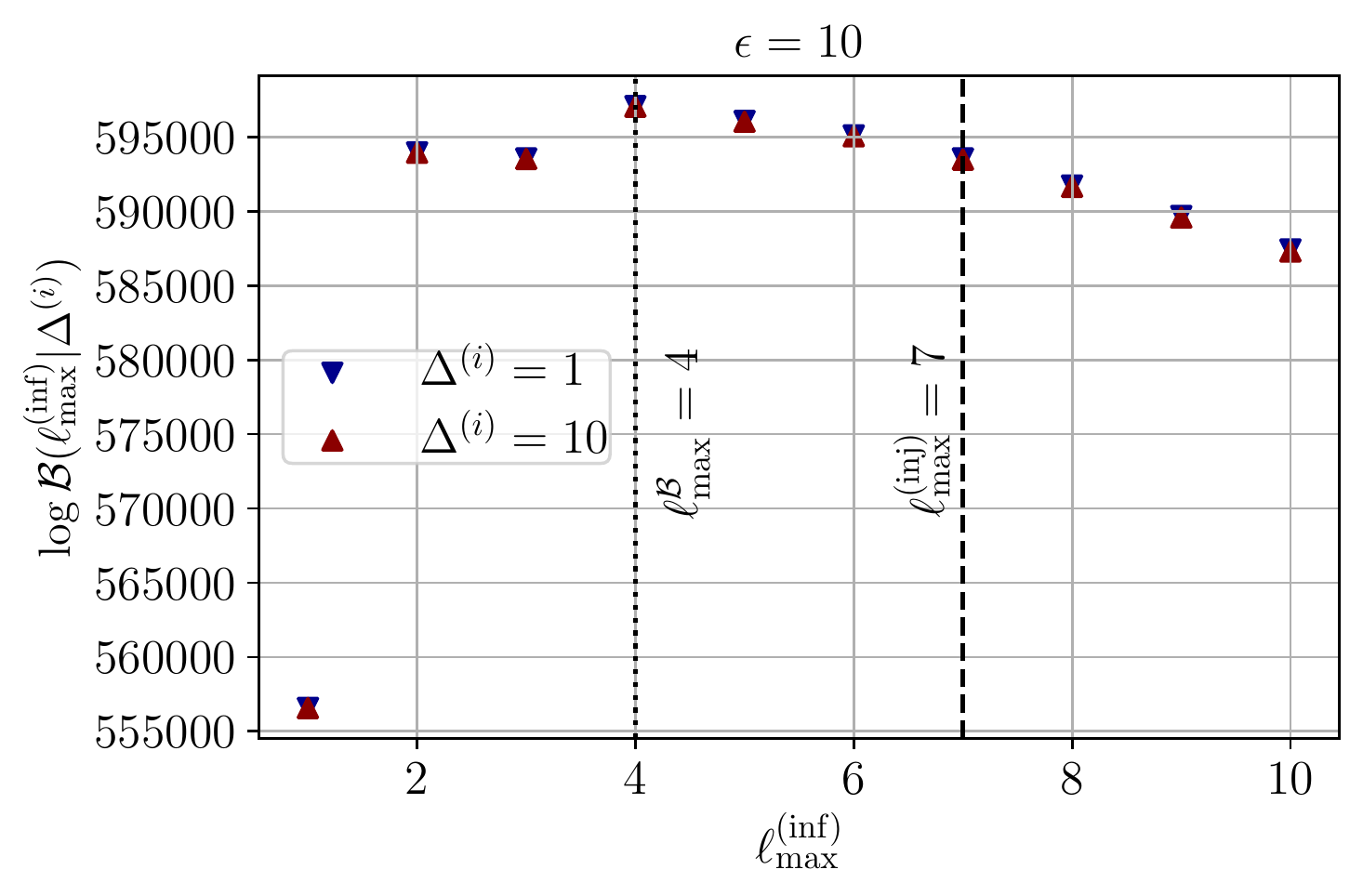}}
\subfloat{\includegraphics[width=0.47\linewidth]{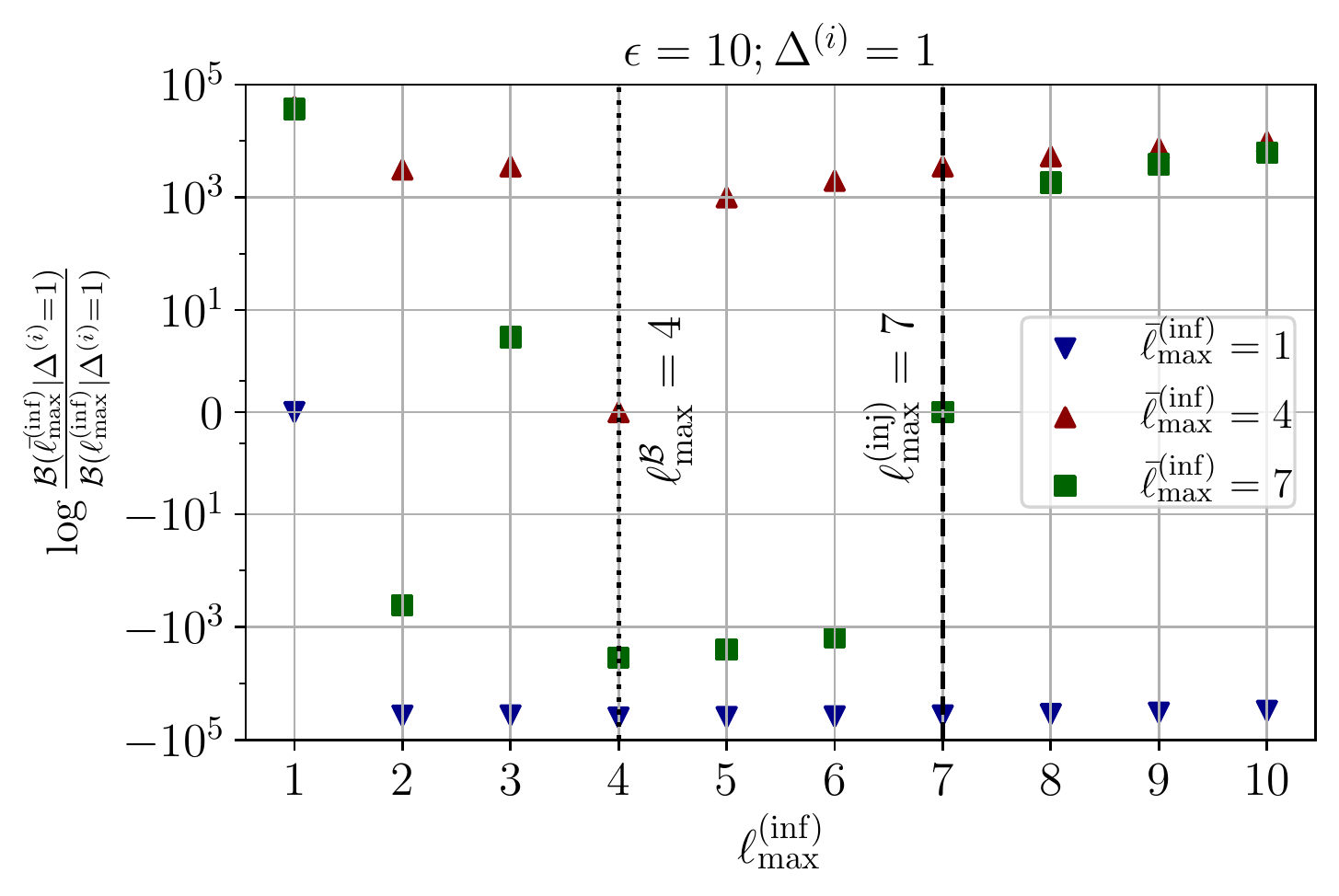}}
\qquad
\subfloat{\includegraphics[width=0.47\linewidth]{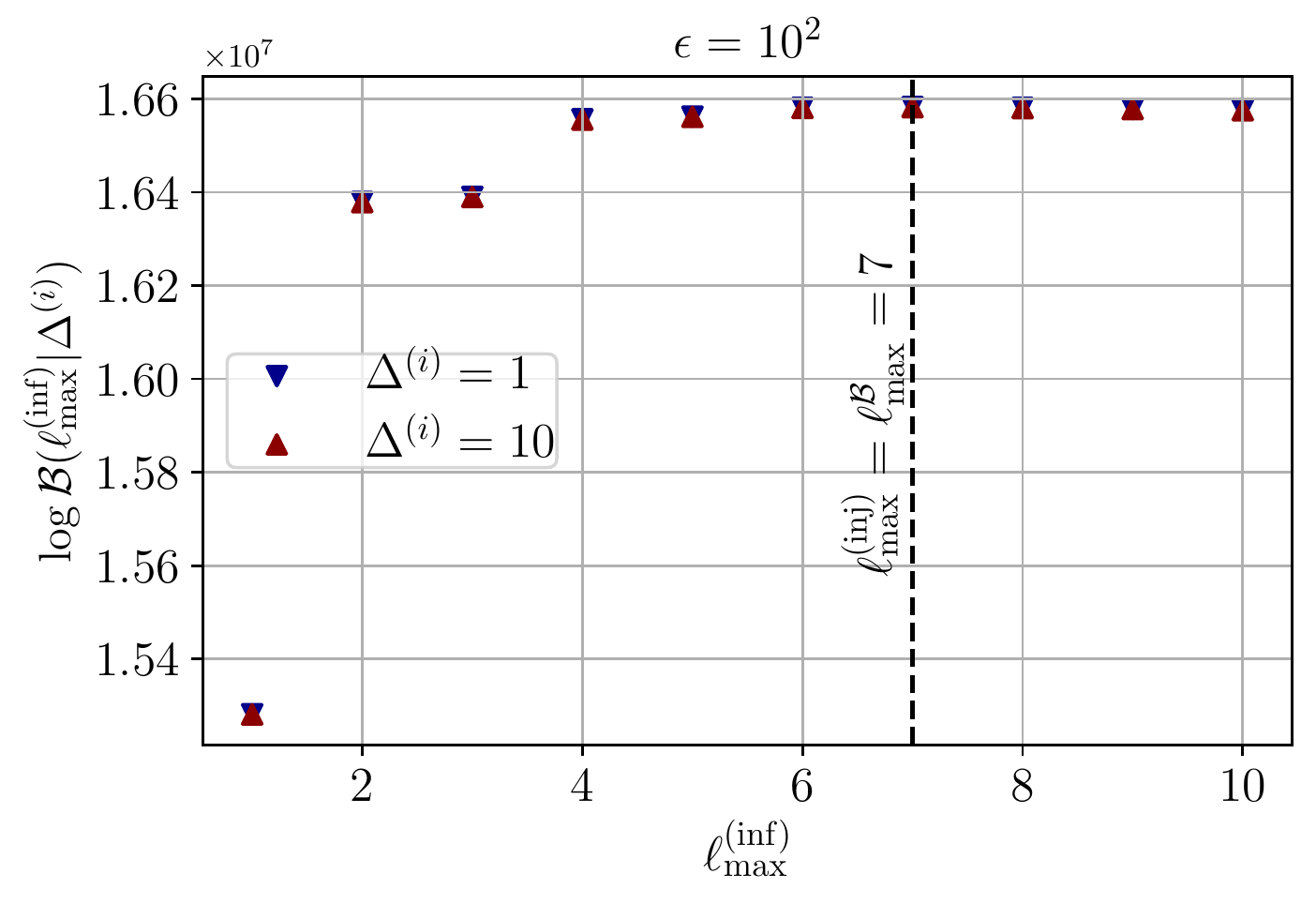}}
\subfloat{\includegraphics[width=0.47\linewidth]{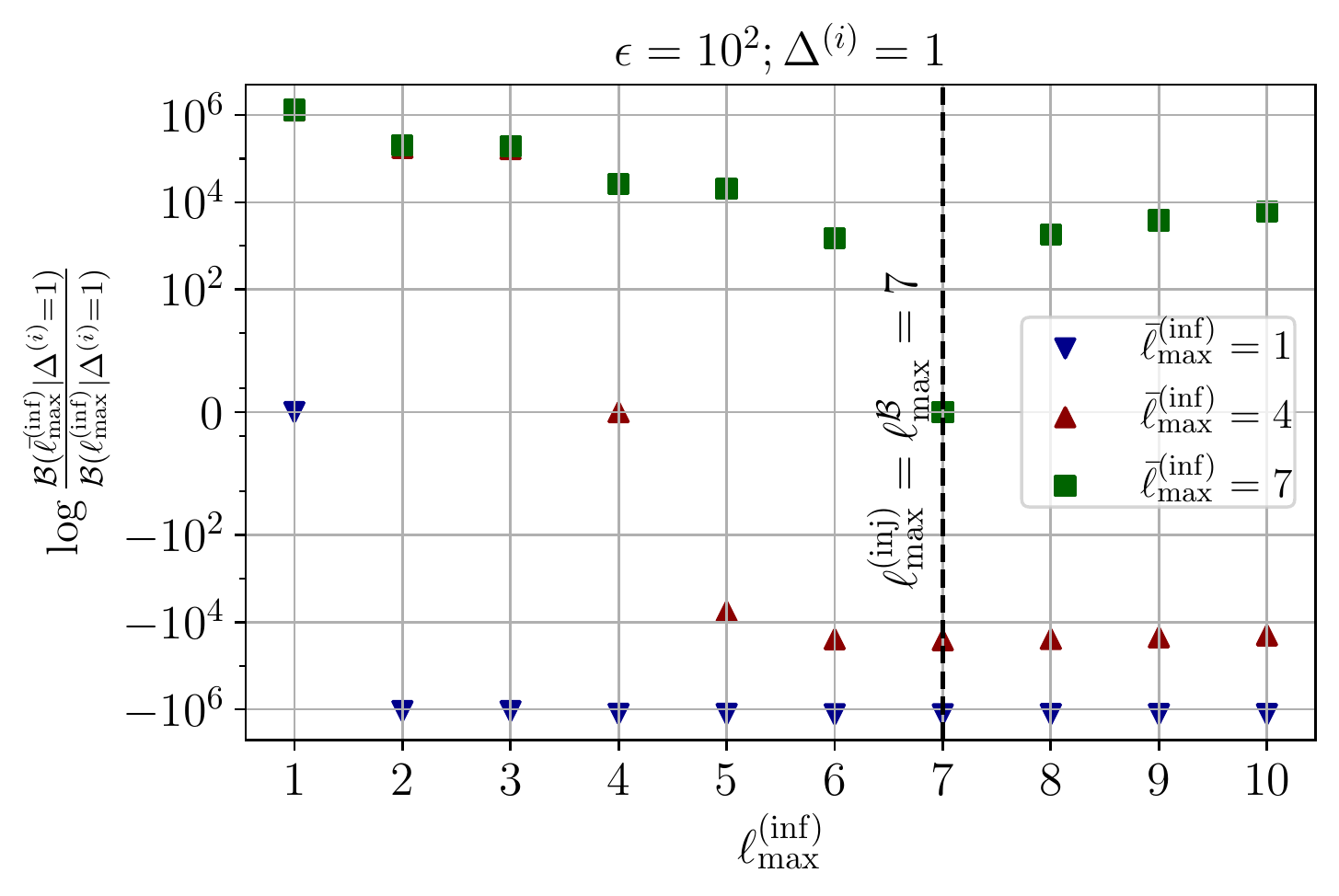}}
\caption{
To rank the hypotheses that we have detected a gravitational-wave background having angular structures up to the angular scale $\ell^{\rm (inf)}_{\rm max}$ ($H_{\ell_{\rm max}^{\rm (inf)}}$) from the data and that the data contain pure noise ($H_{\rm null}$), we compute the Bayes factor between $H_{\ell_{\rm max}^{\rm (inf)}}$ and $H_{\rm null}$ (left panels) and that between $H_{\ell_{\rm max}^{\rm (inf)}}$ and $H_{\bar{\ell}_{\rm max}^{\rm (inf)}}$ (right panels), assuming different widths of the prior ($\Delta^{(i)}$). 
To facilitate the reading of the figures, we represent the maximal angular scale of the simulated background, $\ell_{\rm max}^{\rm (inj)} = 7$, with a dashed vertical line, and the angular scale at which the Bayes factor is maximized, $\ell_{\rm max}^{\mathcal{B}}$, with a dotted vertical line. 
Observe that assuming different $\Delta^{(i)}$ does not significantly affect the resulting logarithm of the Bayes factor, indicating that our analysis is robust against the choice of prior. 
Observe also that as the amplitude of the background increases, as characterized by $\epsilon$, $\ell_{\rm max}^{\mathcal{B}}$ is increasingly consistent with $\ell_{\rm max}^{\rm (inj)}$, until they eventually coincide in the high signal-to-noise ratio scenario. 
This feature is reasonable if we interpret $\ell_{\rm max}^{\mathcal{B}}$ as the maximum resolvable angular scale of the background. 
This pattern suggests that we can determine the angular scale that should be included in the inference analysis by locating the angular scale at which the Bayes factor is maximized, which is consistent with the finding of \cite{Tsukada:2022nsu}. 
}
\label{fig:BayesFactors}
\end{figure*}

Our last mock data analysis concerns the GWB emitted by sources populating the galactic plane. 
For the mock-data challenge of the galactic-plane signal, we focus on $\alpha = 2/3$ because we expect that the results obtained with other choices of $\alpha$ will be quantitatively similar. 
We choose to focus on $\alpha = 2/3 $ because this spectral index corresponds to the background due to CBCs, the only type of GW sources that the Advanced LIGO and Virgo detectors have detected so far. 
To investigate the performance of our analysis when extracting anisotropic GWB signals of different signal-to-noise ratio (SNR, $\rho$), we simulate galactic-plane signals of different SNRs but we reduce the total time length of the mock data of each SNR to 30 days. 
The measurement results from analyzing data of longer time length can be estimated by scaling the SNR, which is proportional to the square root of the integration time.
The $\mathcal{P}_{\ell m}$ of the galactic-plane signal that we simulate are 
\begin{equation}\label{eq:Plm_galactic_plane}
\mathcal{P}^{\rm (inj)}_{\ell m} = \epsilon \mathcal{P}^{\rm (GP)}_{\ell m}, 
\end{equation}
where $\epsilon $ controls the overall amplitude (and SNR) of the galactic-plane signal and $\mathcal{P}^{\rm (GP)}_{\ell m}$ are explicitly given in Appendix~\ref{sec:galactic_plane}.
When choosing these $\mathcal{P}_{\ell m}^{\rm (inj)}$, we set $\ell_{\rm max}^{\rm (inj)} = 7 $ , following \cite{Tsukada:2022nsu}, because this is sufficient to capture the fine angular structures of such a galactic-plane signal. 

The intensity map of the simulated galactic signal is visualized in the top left panel of Fig.~\ref{fig:galactic_plane}, produced using \texttt{HEALPix} \cite{Zonca2019, 2005ApJ...622..759G}. 
The brightness of the color map in all panels represents the intensity, and the intensity of all maps is scaled by a number such that the maximum intensity of each panel is normalized to one. 
The top-right, middle-left, middle-right, bottom-left and bottom-right panels show intensity maps when $\epsilon = 1, 10^{0.5}, 10^{1}, 10^{1.5}, 10^{2}$ respectively, constructed using our analysis, with the spherical harmonic components of the recovered background taken to be the $\mu_i$ of Eq.~(\ref{eq:mu_i}).
As we increase $\epsilon$, the SNR of the signal increases (see the top horizon axis of Fig.~\ref{fig:Match} for the monopole SNR of each $\epsilon$), and the reconstructed intensity map is increasingly consistent with the simulated intensity map. At $\epsilon = 10^2 $, the reconstructed intensity map shows almost no visual differences from the original intensity map. 
The close consistency between the simulated and recovered intensity maps demonstrates the ability of our formalism to resolve detailed and sophisticated angular structures of GWBs. 

Despite the close visual consistency, we also quantitatively assess the consistency between the simulated and reconstructed intensity maps by defining the match, 
\begin{equation}\label{eq:match}
\mathcal{M} = \frac{\sum_{i} w^{\rm (inj)}_i \mu_i}{\sqrt{\sum_{i} \left( w^{\rm (inj)}_i \right)^2}\sqrt{\sum_{i} \mu_i^2}}, 
\end{equation}
where $w^{\rm (inj)}_i$ is the value of the real or imaginary parts of $\mathcal{P}^{\rm (inj)}_{\ell m}$ corresponding to the index $i$, and $\mu_i$ is the recovered value, given by Eq.~(\ref{eq:mu_i}). 
A match closer to unity implies a more faithful recovery. 
If the reconstructed intensity map is identical to the simulated intensity map, $\mathcal{M} = 1$. 
Figure~\ref{fig:Match} shows $\mathcal{M}$ of the simulated galactic-plane signal as a function of $\epsilon$, with the top horizontal axis denoting the SNR of the monopole part of the simulated background of the corresponding $\epsilon$. 
Observe that $\mathcal{M}$ increases to $\sim$ 1 as $\epsilon$ increases. 
This is reasonable because, as the background SNR increases, the angular structures of the simulated anisotropic background can also be more clearly detected. 
Moreover, when the SNR reaches $\sim 400$, which is an SNR that can be achieved within about a year if we detect a GWB of $\Omega_{\rm GW}(f_{\rm Ref}) \sim 10^{-10} $ with the next-generation detectors \cite{2019arXiv190209485H, Dipole_01}, our analysis can recover the intensity map with a match very close to one, indicating its applicability to the realistic detection of a GWB. 

Besides the reconstruction of the intensity map, we also compute the Bayes factor between the hypotheses that there is an anisotropic GWB in the signal and that there is only noise (see Eq.~(\ref{eq:hypothese})), given an injection of an anisotropic GWB from the mock galactic-plane signal.  
The left panels of Fig.~\ref{fig:BayesFactors} show the natural logarithm of the Bayes factor as a function of $\ell_{\rm max}^{\rm (inf)}$, obtained by analyzing the galactic-plane signals of different $\epsilon$, choosing $\Delta^{(i)} = 1$ (inverted blue triangles) and $\Delta^{(i)} = 10$ (red triangles).
Both of these choices of $\Delta^{(i)}$ correspond to a prior of width much larger than the astrophysically motivated value of $w_i$, which should be of $\mathcal{O}(10^{-48})$ (see Fig.~\ref{fig:Noise_results} and \ref{fig:TIDipole_results}). 
The dashed vertical line denotes the $\ell_{\max}^{\rm (inj)}$ of the simulated galactic-plane signal. 
Observe that, in general, for all $\epsilon$, $\log \mathcal{B}(\ell_{\max}^{\rm (inf)}| \Delta^{(i)} = 1)$ is slightly larger than $\mathcal{B}(\ell_{\max}^{\rm (inf)}| \Delta^{(i)} = 10)$ because it has a narrower prior. 
Nonetheless, despite these slight differences, both choices of $\Delta^{(i)}$ lead to a similar Bayes factor. This suggests that, for a reasonably large $\Delta^{(i)}$, the explicit choice of the prior width does not significantly affect the Bayes factor and the hypothesis ranking for the search of anisotropic GWBs.
In this sense, our analysis is robust against different choices of $\Delta^{(i)}$, provided that $\Delta^{(i)}$ is reasonably large. 
Individually, we observe that for a given $\epsilon$, the Bayes factor first increases until it reaches a maximum at a given $\ell_{\rm max}^{\rm (inf)}$, and then it decreases.
Let us denote the $\ell_{\rm max}^{\rm (inf)}$ that maximizes the Bayes factor $\ell_{\rm max}^{\mathcal{B}}$ and show it on Fig.~\ref{fig:BayesFactors} with a dotted vertical line. 
Observe further that $\ell_{\rm max}^{\mathcal{B}}$ depends on $\epsilon$. 
For a larger $\epsilon$, corresponding to a louder signal, $\ell_{\rm max}^{\mathcal{B}}$ is more consistent with $\ell_{\rm max}^{\rm (inf)}$, until eventually $\ell_{\rm max}^{\mathcal{B}}$ coincides with $\ell_{\rm max}^{\rm (inf)}$ in the high signal-to-noise ratio scenario. 
This behavior is reasonable if one interprets the $\ell_{\rm max}^{\mathcal{B}}$ as the maximal resolvable angular scale of the background. 
As we increase $\ell_{\rm max}^{\rm (inf)}$ until $\ell_{\rm max}^{\mathcal{B}}$, we are introducing more parameters in the model that are necessary for a more faithful description of the detectable anisotropic GWB signal. 
More precisely, even if we increase the number of inference parameters, the increase in the marginalized likelihood (the numerator of the Bayes factor) still compensates for the increase in the prior volume. 
Thus, the hypothesis that the detected GWB has nonzero $\mathcal{P}_{\ell_{\rm max}^{\mathcal{B}} m}$ for at least one $m$ between $-\ell_{\rm max}^{\mathcal{B}}$ and $\ell_{\rm max}^{\mathcal{B}}$ is increasingly favored by the data. 
But as we further increase $\ell_{\rm max}^{\rm (inf)}$, the new model parameters are redundant because the detected background shows no resolvable angular structures of the corresponding angular scale. 
The hypothesis that a GWB signal of $\ell_{\rm max}^{\rm (inf)} > \ell_{\rm max}^{\mathcal{B}}$ is detected in the data is now \textit{no longer} better supported by the data than the hypothesis that the signal contains only up to $\ell_{\rm max}^{\mathcal{B}}$, which explains the decrease. 
Finally, if the signal is louder, we can naturally detect the finer angular structures (corresponding to a larger $\ell$) of the simulated background more confidently.
This explains the increasing consistency between $\ell_{\rm max}^{\rm (inf)}$ and $\ell_{\rm max}^{\mathcal{B}}$ as $\epsilon$ increases until $\ell_{\rm max}^{\mathcal{B}}$ essentially coincides with $\ell_{\rm max}^{\rm (inj)}$ in the high SNR limit, when $\epsilon$ is large. 
This behavior could be used to decide which $\ell_{\rm max}^{\rm (inf)}$ is suitable for a particular search, which is also consistent with the discussion in \cite{Tsukada:2022nsu}. 

Apart from competing $H_{\ell^{\rm (inf)}_{\rm max}}$ against $H_{\rm null}$, we can also compete $H_{\ell^{\rm (inf)}_{\rm max}}$ against $H_{\bar{\ell}^{\rm (inf)}_{\rm max}}$, where $\bar{\ell}^{\rm (inf)}_{\rm max}$ is another maximum angular scale included in the inference. 
This can be done by computing the Bayes factor between $H_{\ell_{\rm max}}$ and $H_{\bar{\ell}_{\rm max}}$, which is simply 
\begin{equation}
\mathcal{B}^{\bar{\ell}_{\rm max}^{\rm (inf)}}_{\ell_{\rm max}^{\rm (inf)}} = \frac{p(\{ C\}|H_{\bar{\ell}^{\rm (inf)}_{\rm max}})}{p(\{ C\}|H_{\ell^{\rm (inf)}_{\rm max}})} = \frac{\mathcal{B}(\bar{\ell}^{\rm (inf)}_{\rm max})}{\mathcal{B}(\ell^{\rm (inf)}_{\rm max})}. 
\end{equation}
If $\mathcal{B}^{\bar{\ell}_{\rm max}^{\rm (inf)}}_{\ell_{\rm max}^{\rm (inf)}} > 1$, then $H_{\bar{\ell}^{\rm (inf)}_{\rm max}}$ is favored by the data. 

The right panels of Fig.~\ref{fig:BayesFactors} show $\log \mathcal{B}^{\bar{\ell}_{\rm max}^{\rm (inf)}}_{\ell_{\rm max}^{\rm (inf)}}$ for $\bar{\ell}_{\rm max}^{\rm (inf)} = 1, 4 $, and $7$ as a function of $\ell_{\rm max}^{\rm (inf)}$, obtained by analyzing the galactic-plane signal of $\epsilon =$ 1 (top right), 10 (middle right) and $10^2$ (bottom right). 
We only show the results when $\Delta^{(i)} = 1 $ because the $\log \mathcal{B}^{\bar{\ell}_{\rm max}^{\rm (inf)}}_{\ell_{\rm max}^{\rm (inf)}}$ of $\Delta^{(i)} = 10 $ are qualitatively the same. 
From these panels, we observe the following four patterns in the behavior of $\log \mathcal{B}^{\bar{\ell}_{\rm max}^{\rm (inf)}}_{\ell_{\rm max}^{\rm (inf)}}$ as a function of $\ell_{\rm max}^{\rm (inf)}$:
\begin{enumerate}
    \item $\log \mathcal{B}^{\bar{\ell}_{\rm max}^{\rm (inf)}}_{\ell_{\rm max}^{\rm (inf)}}$ \textit{increases with $\ell_{\rm max}^{\rm (inf)}$}, e.g.~when $\epsilon = 1$, indicating that $H_{\ell_{\rm max}^{\rm (inf)}\leq\bar{\ell}_{\rm max}^{\rm (inf)}}$ is better preferred by the data. 
    \item $\log \mathcal{B}^{\bar{\ell}_{\rm max}^{\rm (inf)}}_{\ell_{\rm max}^{\rm (inf)}}$ \textit{decreases with $\ell_{\rm max}^{\rm (inf)}$}, e.g.~for $\log \mathcal{B}^{\bar{\ell}_{\rm max}^{\rm (inf)} = 1}_{\ell_{\rm max}^{\rm (inf)}}$ when $\epsilon = 10$ and for $\log \mathcal{B}^{\bar{\ell}_{\rm max}^{\rm (inf)} = 1}_{\ell_{\rm max}^{\rm (inf)}}$ and $\log \mathcal{B}^{\bar{\ell}_{\rm max}^{\rm (inf)} = 4}_{\ell_{\rm max}^{\rm (inf)}}$ when $\epsilon = 10^2 $, indicating that $H_{\ell_{\rm max}^{\rm (inf)}\geq\bar{\ell}_{\rm max}^{\rm (inf)}}$ is better preferred by the data. 
    \item $\log \mathcal{B}^{\bar{\ell}_{\rm max}^{\rm (inf)}}_{\ell_{\rm max}^{\rm (inf)}}$ \textit{first decreases, then increases with $\ell_{\rm max}^{\rm (inf)}$ and changes sign at some intermediate $\ell_{\rm max}^{\rm (inf)}$}, e.g.~for $\log \mathcal{B}^{\bar{\ell}_{\rm max}^{\rm (inf)} = 7}_{\ell_{\rm max}^{\rm (inf)}}$ when $\epsilon = 10$, indicating that $H_{\ell_{\rm max}^{\rm (inf)}}$ is preferred over $H_{\bar{\ell}_{\rm max}^{\rm (inf)}}$. 
    In other words, $H_{\bar{\ell}_{\rm max}^{\rm (inf)}}$ is not the hypothesis most preferred by the data. 
    \item $\log \mathcal{B}^{\bar{\ell}_{\rm max}^{\rm (inf)}}_{\ell_{\rm max}^{\rm (inf)}}$ \textit{first decreases, then increases with $\ell_{\rm max}^{\rm (inf)}$ but remains non-negative}, e.g.~for $\log \mathcal{B}^{\bar{\ell}_{\rm max}^{\rm (inf)} = 4}_{\ell_{\rm max}^{\rm (inf)}}$ when $\epsilon = 10$ and for $\log \mathcal{B}^{\bar{\ell}_{\rm max}^{\rm (inf)} = 7}_{\ell_{\rm max}^{\rm (inf)}}$ when $\epsilon = 10^2$, indicating that $H_{\rm \bar{\ell}_{\rm max}^{\rm (inf)}}$ is the hypothesis that is best supported by the data. 
\end{enumerate}
By analyzing these patterns in the behavior of the log Bayes factor ratio shown in the right panels of Fig.~\ref{fig:BayesFactors}, we again conclude that $\mathcal{B}(\ell_{\rm max}^{\rm (inf)})$ peaks at an angular scale that is increasingly consistent with the maximum angular scale contained in the injected background, which is consistent with what we observed from Fig.~\ref{fig:Match} and the left panel of Fig.~\ref{fig:BayesFactors}. 

\section{Concluding remarks}
\label{sec:Conclusion}

In this paper, we presented a novel formalism to analytically marginalize the posterior of the spherical-harmonic components of the intensity map of a GWB in an untargeted Bayesian search. 
By prescribing a wide uniform prior for the real and imaginary parts of the spherical-harmonic components, we approximated the marginalized posterior (or likelihood) and Bayes factor as a Gaussian integral. 
The resulting marginalized posterior is also a Gaussian function. 
By reading off the mean and variance of the marginalized posterior, we can immediately determine the individual maximum posterior value of \textit{many} spherical-harmonic components of the angular distribution of a GWB and gauge the associated measurement uncertainties. 
We validated our formalism by applying it to recover various anisotropic GWBs injections.
For each simulated anisotropic GWB, our analysis accurately extracted the angular structures of the GWB within a $3\sigma$ interval. 
Furthermore, we are able to immediately evaluate the Bayes factor, which is largely unaffected by the width of the uniform prior. 
We showed that the Bayes factor is a reliable indicator of the angular scale that should be included in inference studies in a self-consistent way, which is also consistent with the findings of \cite{Tsukada:2022nsu}. 
As the data products required for our analysis are similar and closely related to those used for existing spherical-harmonic decompositions of the actual data \cite{Thrane:2009fp, LIGOScientific:2011kvu, Romano:2016dpx, LIGOScientific:2016nwa, KAGRA:2021mth, Ain:2018zvo}, we expect that, with minor modifications, our analysis can be applied to actual data to efficiently extract GWB anisotropies along with other existing pipelines. 
Our analysis can also be applied to cross-check the results produced by other existing pipelines that search for anisotropic GWBs. 

Our formalism presents several advantages in the detection of GWBs. 
First, our scheme makes possible Bayesian inference of a larger number of spherical-harmonic components of the angular distribution of a GWB in a reasonable timescale, leading to a much more model-independent Bayesian search of anisotropic GWBs. 
Prior to this work, in principle, we could treat all the spherical-harmonic components of interest as free parameters and attempt to infer them through Bayesian methods, but the computational cost and time needed to numerically sample the posterior would be huge \cite{2019PASA...36...10T}. 
To keep the computational time reasonable, previous Bayesian searches of anisotropic backgrounds either limited the number of spherical-harmonic components inferred (such as in \cite{2021MNRAS.507.5451B}) or precomputed the spherical-harmonic components according to a given model and only inferred the overall amplitude of the anisotropic background (such as in \cite{Tsukada:2022nsu}). 
By analytically marginalizing the posterior, we transform the problem into that of evaluating Gaussian integrals, greatly reducing the time needed to construct the marginalized posterior of spherical-harmonic components and compute the Bayes factor  through Bayesian inference. 
The marginalized posterior of individual spherical-harmonic components can be used to construct an accurate intensity map of the GWB. 
The recovered intensity map can be compared with different GWB models, making the studies of GWBs more efficient.  
Second, our formalism is sufficiently flexible that it can be modified for the search of GWBs in various situations. 
Although this paper lays out the formalism of our method and presents a proof-of-principle analysis of synthetic data, considering only the joint detection of the LIGO Hanford and Livingston detectors, our approach can be straightforwardly extended to a network of detectors.
Moreover, although this paper focused on searching for the GWB of a power-law spectrum, our approach can easily be adapted to the search for anisotropic GWBs of more sophisticated energy densities, such as those described by a broken-power law (such as in \cite{SGWB_PT_02}).

Several aspects of our mock-data analyses differ from those carried out in real searches, but these differences do not undermine the performance of our method when applied to a future search. 
First, in our mock-data analyses, we only considered observations with the LIGO Hanford and Livingston detectors. 
In an actual search, the Virgo detector is operational, and while KAGRA is currently under development, this detector will join the network soon. 
Moreover, next-generation detectors, such as Cosmic Explorer \cite{Reitze:2019iox} and Einstein Telescope \cite{Punturo:2010zz}, are also being planned. 
Our formalism can be easily extended to include these detectors in a future analysis. 
With Virgo and future detectors included, the actual search sensitivity for the detection of a GWB will be greatly improved (assuming the LIGO-Virgo detectors are operating at their design sensitivity), which will also improve the accuracy and performance of our analysis. 
Hence, the results reported in this paper can be regarded as \textit{conservative} estimates of what the future may hold. 
Second, when performing the short-time Fourier transform of the actual data in the time domain, this data will be Hann windowed to avoid spectral leakage \cite{Dipole_01, Thrane:2009fp}. 
To account for the windowing, we need to multiply the mean and variance by windowing factors \cite{LIGOScientific:2003jxj}.  
The full use of the windowed data will then require that any windowed segment has an overlap of 50\% with the Hann window and then be optimally combined.
In this paper, since we are simulating the data in the frequency domain, we did not need to apply these procedures. 
By simulating the data directly in the frequency domain, we are effectively ignoring cross- and auto-correlations due to the serial dependence of the time-domain data. 
However, if the windowing and optimal combinations are correctly implemented, the results of the time-domain analysis should agree well with results that use the likelihood (Eq.~\eqref{eq:likelihood}), which ignores these correlations, as shown in \cite{Dipole_01}. 
Third, the noise we considered in our mock-data challenges was stationary. 
In realistic data, nonstationary and/or non-Gaussian noise transients, also commonly known as ``glitches", may occasionally occur and individual GW signals from CBCs may be present. 
When analyzing the actual data, data segments containing glitches and individual GW signals will be removed upon applying data-quality cuts \cite{O2_directional_notch_list, O3_SHD, LIGOScientific:2016gtq,Thrane:2013npa,Thrane:2014yza,LIGO_data_cut_04}. 
Once these data segments are removed, our formalism can be applied as explained in this paper. 
Fourth, to fully demonstrate the accuracy of resolving the angular structures of GWBs with our method, we assumed strong GWB signals.   
In an actual detection scenario, we expect that GWBs to be much weaker. 
Nonetheless, the signal-to-noise ratio of a GWB detection is approximately proportional to the square root of the detection time \cite{Allen:1996gp, Romano:2016dpx}. 
Thus, in an actual detection, as the integration time is long enough, in principle, we can accumulate a sufficiently large signal-to-noise ratio so that the angular structures of the GWB can be accurately resolved by our analysis. 

Several adaptations or explorations of our method can be carried out in the future to facilitate its implementation and improve its efficiency in the search for anisotropic GWBs in actual data. 
First, when no confident detection of a stochastic background is made, it is insightful to derive the 95\% upper limit on the angular power spectrum, i.e. the 95\% confidence region of 
\begin{equation}\label{eq:C_ell}
C_{\ell} = \left( \frac{2 \pi^2 f_{\rm ref}^3}{3 H_0^2}\right)^2 \frac{1}{2 \ell + 1} \sum_{m = - \ell}^{\ell} \left[|\mathcal{P}^{\rm Re}_{\ell m}|^2 + |\mathcal{P}^{\rm Im}_{\ell m}|^2\right]. 
\end{equation}
Since the individual $\mathcal{P}^{\rm Re}_{\ell m}$ and $\mathcal{P}^{\rm Im}_{\ell m}$ follow a Gaussian marginalized posterior whose mean is nonzero in general, as shown by our calculations, $C_{\ell}$ follows a \textit{generalized} chi-squared distribution, which does not admit a simple closed-form analytic expression for its cumulative probability distribution function. 
Instead, numerical means are still required for constructing the cumulative probability distribution function of a generalized chi-squared distribution. 
Further effort must be devoted to either derive analytic results or to develop efficient numerical schemes that rapidly reconstruct the upper limit on $C_{\ell} $ when there is no GWB detection. 

Second, to analytically marginalize the posterior, we prescribe a wide prior for the spherical harmonic components. 
Within the prior space, some spherical harmonic components actually correspond to an intensity map of negative intensity along some sky directions, which is not physical. 
Prescribing wide priors also makes our analysis sub-optimal, in the sense that it may need much higher SNR to detect or resolve the angular structure of a GWB. 
One possible way to improve the method is to prescribe a conjugate normal prior for the spherical-harmonic components, which does not require the prior to be wide, thereby reducing the prior space that corresponds to negative intensity. 
However, the marginalized posterior and Bayes factor will then depend on the properties of the conjugate normal prior. 
Another possible way to improve the method is to make use of Clebsch-Gordan coefficients to parameterize the intensity map of a GWB \cite{2021MNRAS.507.5451B}. 
However, the exponent of the likelihood in terms of Clebsch-Gordan coefficients becomes quartic in the relevant parameters. 
The analytical marginalization of such a posterior may be possible through an appropriate change of variables, but this requires further exploration. 

Third, our analysis uses a spherical harmonic basis, which is well-suited for the search of wide-spread GWB sources. 
However, point-like sources, such as nearby galaxy superclusters, may also contribute to anisotropic GWBs. 
These sources can be more adequately described using the pixel basis \cite{Ballmer:2005uw}. 
To include these point-like sources in our search, we should explore extending our work to incorporate such a basis. 
Working with the pixel basis may require many more parameters to characterize GWB anisotropies than the spherical-harmonic basis. 
Therefore, in future work, one could explore how to perform the analysis with the pixel basis within a reasonable time frame.

Fourth, the marginalization of the likelihood in joint inferences of a GWB and individually resolvable GW signals requires further investigation. 
As mentioned here and also pointed out by \cite{2021MNRAS.507.5451B}, a motivation to measure the angular structure of GWBs in a Bayesian way is its integration with the existing search of other GW signals, such the those emitted by CBCs.
One formalism that is capable of simultaneously searching for GWBs and individual GW signals is the ``master-likelihood'' method (also known as the hyper-likelihood approach) \cite{master_likelihood_01, master_likelihood_02}. 
The marginalization of the master likelihood over the spherical harmonic components is certainly worth exploring to unite the search approaches of different types of GW signals for search efficiency reasons.  

Finally, our formalism essentially assumes that we are searching for stationary GWBs. 
However, the kinematic dipole of a GWB induced by the proper motion of the Earth around the Solar System barycenter, a guaranteed anisotropic signal of GWBs \cite{Allen:1996gp, Dipole_02, Dipole_03}, is time-dependent and requires a specially targeted method to implement in a search \cite{Dipole_01}. 
As this type of GWB signal varies over a timescale that is much longer than a sidereal day, we expect that our formalism can be straightforwardly adapted, say, by including this mild time dependence of the signal into the likelihood (E.q.~(\ref{eq:likelihood})) before marginalization, to search for these GWB signals. 
Nonetheless, more exploration is still needed to determine the optimal way to modify our formalism to search for GWB signals with time dependence. 

\section*{Acknowledgements}

The authors would like to thank Erik Floden, Vuk Mandic, Joesph Romano and Leo Tsukada for insightful discussion, Sharan Banagiri, Sanjit Mitra, and Joesph Romano for providing the spherical-harmonic components of the galactic plane for the mock data analyses, and Neil Cornish, Alexander Jenkins, Xavier Siemens, and Leo Tsukada for comments on the initial manuscript. 
N.Y. and A.C. acknowledge support from the Simmons Foundation through Award No. 896696 and the NSF through award PHY-2207650.
The numerical results reported in this paper were produced using the workstation of the CUHK GW working group and the Illinois Campus Cluster, a computing resource that is operated by the Illinois Campus Cluster Program (ICCP) in conjunction with National Center for Supercomputing Applications (NCSA), and is supported by funds from the University of Illinois at Urbana-Champaign.

\appendix

\section{$\mathcal{P}_{\ell m}$ of the galactic-plane signal injection}
\label{sec:galactic_plane}

Below we provide the $\mathcal{P}_{\ell m}$ for the mock galactic plane signals that we simulated. 
The $\mathcal{P}_{\ell m}$ numbers are stored by $m$, in accordance with the convention of \texttt{HEALPix}. 
The relative intensity map is not alternated if one scales all $\mathcal{P}_{\ell m}$ by the same constant. 

\begin{align*}
\mathcal{P}^{\rm (GP)}_{00} & = 6.24\times 10^{-48}, \\ 
\mathcal{P}^{\rm (GP)}_{10} & =-1.92\times 10^{-50}, \\
\mathcal{P}^{\rm (GP)}_{20} & =1.28\times 10^{-49}, \\
\mathcal{P}^{\rm (GP)}_{30} & = -1.78\times 10^{-49}, \\
\mathcal{P}^{\rm (GP)}_{40} & = -1.03\times 10^{-49}, \\
\mathcal{P}^{\rm (GP)}_{50} & = -8.89\times 10^{-50}, \\
\mathcal{P}^{\rm (GP)}_{60} & = -3.63\times 10^{-49}, \\
\mathcal{P}^{\rm (GP)}_{70} & = -4.82\times 10^{-50}, \\
\mathcal{P}^{\rm (GP)}_{11} & = -2.90\times 10^{-52}-5.54\times 10^{-50}i, \\
\mathcal{P}^{\rm (GP)}_{21} & =  -9.05\times 10^{-49}-1.20\times 10^{-49}i, \\
\mathcal{P}^{\rm (GP)}_{31} & =  -1.39\times 10^{-50}+2.90\times 10^{-49}i, \\
\mathcal{P}^{\rm (GP)}_{41} & = -5.72\times 10^{-50}+9.50 \times 10^{-50}i, \\
\mathcal{P}^{\rm (GP)}_{51} & = 6.89\times 10^{-51}-6.96\times 10^{-50}i, \\
\mathcal{P}^{\rm (GP)}_{61} & = 3.86\times 10^{-51}-4.58\times 10^{-50}i, \\
\mathcal{P}^{\rm (GP)}_{71} & = -2.10\times 10^{-50}+3.32\times 10^{-50}i, \\
\mathcal{P}^{\rm (GP)}_{22} & = -9.95\times 10^{-49}-2.92\times 10^{-49}i, \\
\mathcal{P}^{\rm (GP)}_{32} & = -8.70\times 10^{-50}-1.36\times 10^{-49}i, \\
\mathcal{P}^{\rm (GP)}_{42} & = 3.92\times 10^{-49}+8.90 \times 10^{-50}i, \\
\mathcal{P}^{\rm (GP)}_{52} & = 7.52\times 10^{-50}-2.21\times 10^{-49}i, \\
\mathcal{P}^{\rm (GP)}_{62} & = 3.31\times 10^{-49}-2.03\times 10^{-50}i, \\
\mathcal{P}^{\rm (GP)}_{72} & = 3.46\times 10^{-50}-5.64\times 10^{-50}i, \\
\mathcal{P}^{\rm (GP)}_{33} & = -1.85\times 10^{-50}+1.53\times 10^{-50}i, \\
\mathcal{P}^{\rm (GP)}_{43} & = 5.59\times 10^{-49}+9.57\times 10^{-50}i, \\
\mathcal{P}^{\rm (GP)}_{53} & = 1.14\times 10^{-50}-1.19\times 10^{-50}i, \\
\mathcal{P}^{\rm (GP)}_{63} & = 1.69\times 10^{-49}-1.29\times 10^{-50}i, \\
\mathcal{P}^{\rm (GP)}_{73} & = -4.30\times 10^{-50}+1.01\times 10^{-49}i, \\
\mathcal{P}^{\rm (GP)}_{44} & = 3.46\times 10^{-49}+2.89\times 10^{-49}i, \\
\mathcal{P}^{\rm (GP)}_{54} & = 1.00\times 10^{-49}+1.47\times 10^{-49}i, \\
\mathcal{P}^{\rm (GP)}_{64} & = -2.59\times 10^{-49}-1.33\times 10^{-49}i, \\
\mathcal{P}^{\rm (GP)}_{74} & = -1.68\times 10^{-50}+5.37\times 10^{-50}i, \\
\mathcal{P}^{\rm (GP)}_{55} & = 1.29\times 10^{-50}+1.42\times 10^{-49}i, \\
\mathcal{P}^{\rm (GP)}_{65} & = -3.35\times 10^{-49}-1.58\times 10^{-49}i, \\
\mathcal{P}^{\rm (GP)}_{75} & = -7.04\times 10^{-50}-4.37\times 10^{-51}i, \\
\mathcal{P}^{\rm (GP)}_{66} & = -1.66\times 10^{-48}-1.51\times 10^{-49}i, \\
\mathcal{P}^{\rm (GP)}_{67} & = 3.05\times 10^{-50}-1.95\times 10^{-49}i, \\
\mathcal{P}^{\rm (GP)}_{77} & = -5.54\times 10^{-51}-1.72\times 10^{-49}i \\
\end{align*} 

\bibliography{ref}

\end{document}